%% file: mj-clustering_2_PAPERv3.tex
\documentclass[preprint,nofootinbib,aps,superscriptaddress,reprint,floatfix]{revtex4}
\usepackage{etoolbox}
\makeatletter
\patchcmd{\maketitle}{\@fpheader}{}{}{}
\makeatother

\usepackage{graphicx,epsf,bm}
\usepackage{amssymb}
\usepackage{amsmath}
\usepackage{epstopdf}
\usepackage{booktabs}
\usepackage{subfigure}
\usepackage{rotating}
\usepackage{xspace}
\usepackage{slashed}

\newcommand{\pythia}{\textsc{Pythia}\xspace}

\newcommand{\madgraphamc}{\textsc{MadGraph5\_aMC@NLO}\xspace}
\newcommand{\delphes}{\textsc{Delphes}\xspace}
\newcommand{\fastjet}{\textsc{FastJet}\xspace}

\newcommand{\GeV}{\ensuremath{\,\text{GeV}}}
\newcommand{\TeV}{\ensuremath{\,\text{TeV}}}
\newcommand{\invfb}{\ensuremath{\,\text{fb}^{-1}}}

\usepackage{color}

\begin{document}

\title{\bf \Large Jumping into buckets,\\ or How to decontaminate overlapping fat jets}

\author{\large  Koichi Hamaguchi}
\affiliation{Department of Physics, University of Tokyo, Bunkyo-ku, Tokyo 113--0033, Japan.}
\affiliation{Kavli Institute for the Physics and Mathematics of the Universe (Kavli IPMU), \\
University of Tokyo, Kashiwa 277--8583, Japan}

\author{\large  Seng Pei Liew}
\affiliation{Department of Physics, University of Tokyo, Bunkyo-ku, Tokyo 113--0033, Japan.}

\author{\large  Martin Stoll}
\affiliation{Department of Physics, University of Tokyo, Bunkyo-ku, Tokyo 113--0033, Japan.}

\begin{abstract}
At the LHC, tagging boosted heavy particle resonances which decay hadronically,
such as top quarks and Higgs bosons,
can play an essential role in new physics searches.
In events with high multiplicity, however, the standard approach to tag boosted resonances by a large-radius fat jet becomes difficult 
because the resonances are not well-separated from other hard radiation.
In this paper, we propose a different approach to tag and reconstruct boosted resonances by using the recently proposed mass-jump jet algorithm.
A key feature of the algorithm is the flexible radius of the jets, 
which results from a terminating veto that prevents the recombination of two hard prongs %with a substantial mass jump.
if their combined jet mass is substantially larger than the masses of the separate prongs.
The idea of collecting jets in ``buckets" is also used.
As an example, we consider the fully hadronic final state of pair-produced vectorlike top partners at the LHC, $pp\to T\bar{T}\to t\bar{t}HH$,
and show that the new approach works well.
We also show that tagging and kinematic reconstruction of boosted top quarks and Higgs bosons are possible with good quality even in these very busy final states.
The vectorlike top partners are kinematically reconstructed, which allows their direct mass measurement.

\end{abstract}

\preprint{UT-15-17}
\preprint{IPMU-15-0068}

\def\thepage{{}}
\maketitle
\def\thepage{\arabic{page}}

\section{Introduction}
\label{sec:intro}

The high-energy frontier of particle physics is often probed at  hadronic colliders,
such as Tevatron and the Large Hadron Collider (LHC).
Most of the time, collisions at hadron colliders produce coloured partons that undergo showering and hadronization resulting in a large number of final-state hadrons. In order to extract momentum, energy, and other information of the hard-scattering partons, it is necessary to cluster the hadrons observed at the detectors into jets.
Studying jets and their clustering procedure (jet algorithm) are therefore the keys to understanding the physics at hadron colliders.    

It is particularly interesting to consider the clustering of hadrons originating from boosted heavy particle resonances such as top quarks and Higgs bosons, 
given that the LHC restarts with an increased energy $\sqrt{s}=13$-$14\TeV$ and will 
produce such resonances copiously. The construction of large-radius ``fat'' jets has become a common approach to deal with such scenarios, facilitated by progress in tagging of boosted resonances (see e.g.~Refs.~\cite{Altheimer:2012mn,Adams:2015hiv} and references therein). 
By investigating the substructure of a fat jet, more information on the energy deposit pattern is available compared to separately resolved small-radius jets.
Furthermore, the classical problem of finding an optimal 
jet radius~\cite{Dasgupta:2007wa,Cacciari:2008gd,Soyez:2010rg} is avoided and jet combinatorics are significantly reduced.

\begin{figure}[ht]
 \begin{center}
  \includegraphics[width=0.7\textwidth]{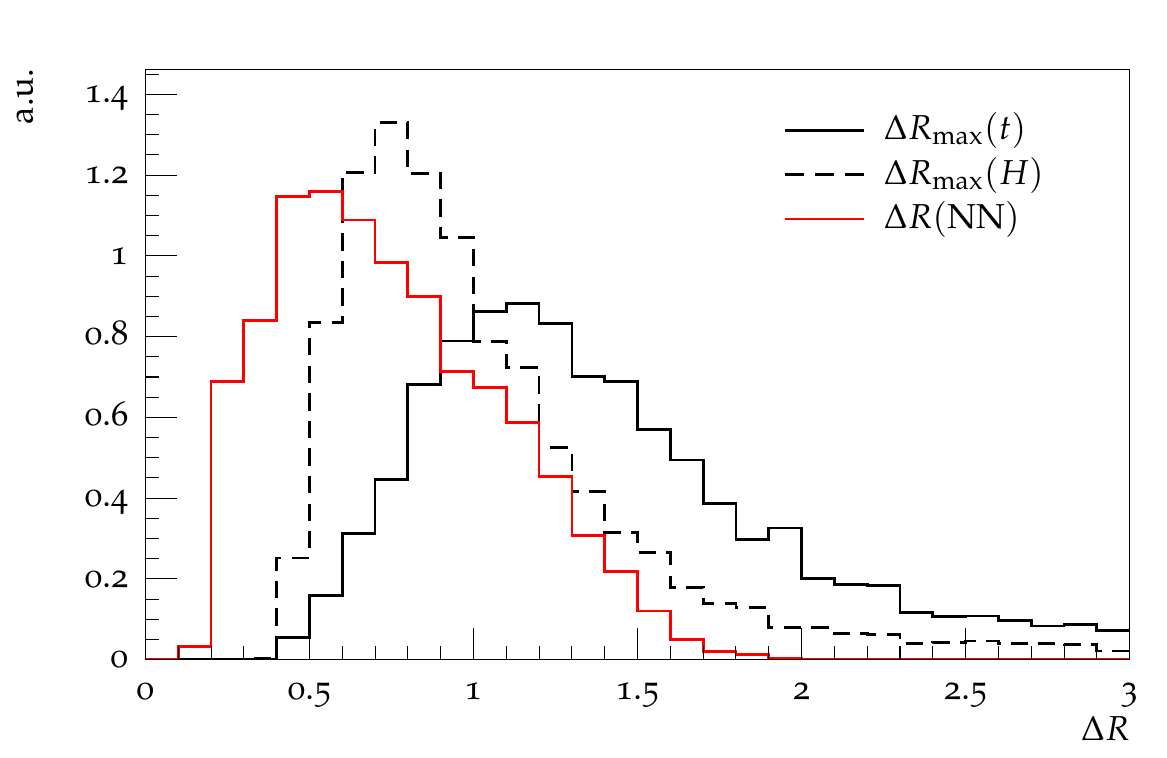}
 \end{center}
 \caption{
 Angular distances between daughter particles from top and Higgs decays,
  in the benchmark process  $pp\to T\bar{T}\to t\bar{t}HH\to 10\text{ jets}$
  at the LHC with $\sqrt{s}=14\TeV$, with vectorlike top mass $m_T=1\TeV$ (parton level, arbitrary units).
 The largest $R$-distance between the daughters of the same top quark (Higgs boson) is denoted by $\Delta R_\text{max}(t)$ ($\Delta R_\text{max}(H)$) and plotted with black solid (dashed) lines.
 The minimal $R$-distance between the nearest neighbour daughters coming from different mothers $\Delta R(\text{NN})$ is depicted by the red line. 
See Sec.~\ref{sec:analysis} for details.
 }
 \label{fig:contamination_deltar}
\end{figure}

A more challenging situation arises when multiple heavy resonances are produced simultaneously.
Such processes lead to very busy final states where the heavy particles under consideration (such as top and Higgs) as well as their daughter particles are not well-separated.
(See Fig.~\ref{fig:contamination_deltar}.)
As a result, fat jets merge in most events, and a majority of the jets contain decay products from more than one resonance.
Such a scenario is not adequately addressed by most tagging algorithms based on (isolated) fat jets.

In the present work, we suggest a new framework of jet tagging that allows particle reconstruction with good quality compared to traditional methods in such a busy hadronic environment. A key ingredient is the ``mass-jump'' jet clustering algorithm~\cite{Stoll:2014hsa}, which is an extension of the ``mass-drop'' subjet identification in fat jets as employed in the HEPTopTagger~\cite{1006.2833}.
It has been shown that the mass-jump algorithm 
gives competitive performance, but now the ``sub''jets are formed directly without the definition of an intermediate fat jet.
Mass-jump clustering harnesses the advantage of fat jet substructure algorithms of resolving small jets without reference to a fixed radius.

The absence of the fat jet, however, reintroduces the issue of large combinatorics in such busy environments,
since the decay products of the heavy resonances cannot be disentangled a priori.
To facilitate event reconstruction, the idea of collecting jets into separate ``buckets"~\cite{1302.6238,1310.6034} is applied, which allows efficient assignment of jets to their respective resonances.

The jet-tagging method proposed here is applicable to a broad range of Standard Model (SM) and beyond the SM phenomena at hadron colliders. There are indeed important  SM processes which involve decays of multiple heavy particles, resulting in a busy hadronic environment. A prime example is the associated production of a Higgs boson with two top quarks ($pp \to t\bar{t}H$), which has attracted attention as this channel opens up the opportunity to measure directly the Higgs-top Yukawa coupling, an essential probe towards understanding the Higgs sector.

Some models of supersymmetry (SUSY) also predict large multiplicity of jets with little or no missing transverse energy (MET). For example, assuming that the gluino is the lightest SUSY particle, it can decay into a top quark and jets when baryonic R-parity associated to the third generation quark is violated~\cite{Barbier:2004ez}. This leads to a multijet final state when the top quark decays hadronically ($\tilde{g}\tilde{g}\to t t j j j j$). Another example is the stealth SUSY, where the top and the lighter stop ($\tilde t_1$) masses are almost degenerate, leading to final states without significant MET~\cite{Fan:2011yu}. The heavier stop ($\tilde t_2$) has a model-dependent decay branching ratio to the Z or Higgs boson (${\tilde t_2} \to {\tilde t_1} Z/H$). The hadronic mode of this decay again leads to multijet final states with little MET ($\tilde{t_2}\tilde{t_2^*}\to t \bar{t} (H/Z)(H/Z)$). Our method may allow to fully reconstruct the underlying new particles in such models, too.

In order to illustrate the strength of our jet tagging method, we investigate a simplified version of heavy particle production topology in this paper, i.e.~we consider the fully hadronic final state of pair-produced vectorlike top partners at the LHC ($pp \to T\bar{T} \to t\bar{t}HH$). In particular, we study the performance of our taggers at the 14 TeV LHC with a vectorlike top of mass around 1 TeV as our benchmark scenario.
Studies of fully hadronic final states in similar processes have been based on fat jet substructure~\cite{Gopalakrishna:2013hua,Girdhar:2014wua,Yang:2014usa,Endo:2014bsa}, including experimental searches at 8 TeV by CMS~\cite{Khachatryan:2015axa}.
Current exclusion bounds on the vectorlike top mass are $m_T \gtrsim 700 - 950$ from ATLAS~\cite{Aad:2014efa,Aad:2015mba,ATLAS-CONF-2015-012} and $m_T \gtrsim 690 - 910$ from CMS~\cite{Chatrchyan:2011ay,Chatrchyan:2013uxa,CMS:2014dka}, depending on the assumed branching fractions.

This paper is arranged as follows:
In Section~\ref{sec:recap}, we review the essential tools used in our analysis:
the mass-jump clustering and the bucket algorithms.
We then apply our method to the benchmark scenario of the fully hadronic decay of pair-produced vectorlike tops in Section~\ref{sec:analysis}.
The performance of the involved top and Higgs taggers is investigated  in Section~\ref{sec:tagging}.
We conclude our findings in Section~\ref{sec:summary}.

\section{Recap: mass-jump jet clustering and buckets}
\label{sec:recap}

In this paper, we investigate a new approach of analyzing high-multiplicity final states based on separately resolved jets. 
We try to answer the two key questions that arise in such an analysis:
(1) which algorithm to use to construct the jets, and
(2) how to reduce the sheer combinatorial choices of assigning the jets to the resonance particles of the process.
We examine the first question by comparing the mass-jump algorithm~\cite{Stoll:2014hsa}
with the corresponding jet clustering algorithm of the generalized $k_T$ family.
The latter question is addressed by the bucket algorithm introduced in Refs.~\cite{1302.6238,1310.6034}.
Both recent techniques are briefly reviewed in the remainder of this section.

\subsection{Jet clustering with a mass-jump veto}
\label{sec:recap:mj}

In the commonly used generalized-$k_T$ family of jet clustering algorithms, jets are constructed by sequential recombination of input particles until a certain angular distance is reached, the jet radius $R$.
As a result, all jet centres are mutually separated by $\Delta R(j_1,j_2)=\sqrt{(\phi_1-\phi_2)^2+(y_1+y_2)^2} > R$ and the angular spread of each jet is roughly $\leq R$.
$\phi$ and $y$ are the azimuthal angle and rapidity, 
respectively.
The choice of a large parameter $R$ can lead to radiation from two (or more) hard partons ending up in the same jet (splash-in).
On the other hand, if the radius is too small, not all final-state radiation of a hard parton will be captured by the jet (splash-out).
In both cases, jet-parton correspondence is disturbed. 

The mass-jump clustering algorithm~\cite{Stoll:2014hsa} addresses this problem by implementing a flexible jet radius based on an intrinsic jet property (jet mass) as well as the topology of jets in its vicinity.
Two parameters are introduced accordingly, the jet mass threshold $\mu$ and the mass-ratio parameter $\theta$.
Jet clustering starts from a set of input particles, which are labelled {\it active} protojets. A distance metric for protojets $j_i$ is defined as
\begin{align}
   d_{j_1j_2} = \frac{\Delta R_{j_1j_2}^2}{R^2} \min\left[p_{j_1\perp}^{2n},p_{j_2\perp}^{2n}\right] \,,\quad
   d_{j_1B} = p_{j_1\perp}^{2n} \,,
   \label{eq:jet_metric}
\end{align}
where $n=1$ corresponds to the $k_T$ algorithm~\cite{Catani:1991hj,Catani:1993hr,Ellis:1993tq}, $n=0$ to the Cambridge/Aachen algorithm~\cite{Dokshitzer:1997in,Wobisch:1998wt}, and $n=-1$ to the anti-$k_T$ algorithm~\cite{Cacciari:2008gp}.
The sequential recombination algorithm then proceeds as follows~\cite{Stoll:2014hsa}:

  \begin{enumerate}
   \item Find the smallest  $d_{j_aj_b}$ among active protojets, including $d_{j_aB}$; if it is given by a beam distance, $d_{j_aB}$, label $j_a$ {\it passive} and repeat step 1.
      \item Combine $j_a$ and $j_b$ by summing their four-momenta, $p_{j_a+j_b}=p_{j_a}+p_{j_b}$ ($E$-scheme).
   If the new jet is still light, $m_{j_a+j_b}<\mu$, replace $j_a$ and $j_b$ by their combination in the set of active protojets and go back to step 1.\\
   Otherwise check the mass-jump criterion: If $\theta\cdot m_{j_a+j_b} >  \max\left[m_{j_a},m_{j_b}\right]$ label $j_a$ and $j_b$ {\it passive} and go back to step 1.
   
   \item Mass jumps can also appear between an active and a passive protojet.
   To examine this
   \begin{itemize}
    \item[a.] Find the passive protojet $j_n$ which is closest to $j_a$ in terms of the metric $d$ and is not isolated, $d_{j_aj_n}<d_{j_nB}$.
    \item[b.] Then check if these two protojets would have been recombined if $j_n$ had not been rendered passive by a previous veto, i.e.~$d_{j_aj_n}<d_{j_aj_b}$.
    \item[c.] Finally check the mass-jump criterion, $m_{j_a+j_n}\geq\mu$ and $\theta\cdot m_{j_a+j_n}>\max\left[m_{j_a},m_{j_n}\right]$.
   \end{itemize}
   If all these criteria for the veto are fulfilled, label $j_a$ {\it passive}.
   Do the same for $j_b$.
   If either of $j_a$ or $j_b$ turned passive, go back to step 1.   
   \item No mass-jump has been found, so replace $j_a$ and $j_b$ by their combination in the set of active protojets. Go back to step 1.
  \end{enumerate}
  Clustering terminates when there are no more active protojets left. Passive protojets are then labelled jets.
  Note that for $\theta=0$ or $\mu=\infty$ standard sequential clustering without veto is recovered, which is reduced to steps 1 and 4.
Jet clustering can be $k_T$-like, C/A-like, or anti-$k_T$-like, depending on the metric chosen [see Eq.~(\ref{eq:jet_metric})].
  
In Ref.~\cite{Stoll:2014hsa} it has been shown that the mass-jump clustering algorithm can be useful to resolve the close-by subjets of boosted top quarks.
At the same time, isolated jets are hardly affected by the veto if $\mu$ and $\theta$ are not chosen too aggressively.
These properties qualify the mass-jump algorithm as a suitable candidate for processes with very busy final states where this flexibility is essential.

The mass-jump algorithm described above is the first member of the family of jet clustering algorithms with a terminating veto and has been made publicly available as part of the \fastjet contribution package~\cite{fastjet-contrib}.
The plugin is dubbed \texttt{ClusteringVetoPlugin} and accepts any user-defined veto function.
Its exemplary usage is illustrated within the package.

\subsection{The bucket algorithm}

In high-multiplicity events, the assignment of jets to their respective heavy resonances can easily get out of hand.
There are $6!/(3!3!)=20$ possible combinations to assign six jets to two top quarks, and for eight jets coming from two tops and a Higgs boson this number already reaches $8!/(3!3!2!)=560$.
The bucket algorithm~\cite{1302.6238,1310.6034} was proposed in the context of these two final states and introduces a ``bucket'' for each top quark (and one additional bucket of unassigned jets $B_\text{ISR}$), into which the jets are allocated.
For a bucket $B_i$ the metric
\begin{align}
 \Delta_{B_i} = |m_{B_i}-m_t| \qquad\text{with}\qquad
 m_{B_i}^2 = \left(\sum_{j\in B_i} p_j\right)^2
\end{align}
measures the similarity of a collection of jets inside the bucket with a top quark.
In Refs.~\cite{1302.6238,1310.6034}, the combination  
is determined by minimizing a global $\chi^2$-like metric defined as
\begin{align}
 \Delta^2 = \omega \Delta_{B_1}^2 + \Delta_{B_2}^2 \,,
\end{align}
and choosing a large $\omega=100$ effectively decouples the two buckets.
Thereby $\Delta_{B_1}<\Delta_{B_2}$ holds and the problem of unfeasible combinatorics is circumvented because the buckets can be filled independently.
Top tagging is performed by imposing cuts on each bucket later on.
Whereas in the original proposal the number of jets inside each bucket is not fixed, in this paper we require strictly three jets in each top bucket, and also introduce Higgs buckets that contain exactly two jets.\footnote{In the top quark rest frame, in a large fraction of events, one of the decay products from $t\to bW^+\to bjj$ carries low transverse momentum and thus fails to be reconstructed as a jet. As this paper is concerned with boosted top quarks from a heavy resonance decay, this problem does not occur.}

We apply the bucket algorithm in conjunction with mass-jump jet clustering as well as conventional jet clustering for comparison.
Our benchmark scenario is given by $t\bar{t}HH$ production from a pair of vectorlike tops.
Clearly the naive combinatorics are overwhelming even for the minimal final-state multiplicity of ten jets: $10!/(3!3!2!2!)=25200$.
To tackle this problem we formally define a global metric
\begin{align}
 \Delta^2 = \omega_1 \Delta_{B_{t1}}^2 + \omega_2 \Delta_{B_{t2}}^2 +
            \omega_3 \Delta_{B_{H1}}^2 + \omega_4 \Delta_{B_{H2}}^2
 \label{eq:bucket_global_metric}
\end{align}
and explicitly decouple the four buckets by choosing the (positive) weights such that
\begin{align}
 \frac{\omega_{i+1}}{\omega_i} = +0 \quad i=1..3 \,.
 \label{eq:bucket_decoupled_metric}
\end{align}
Therefore, the buckets are filled separately in order $(B_{t1},B_{t2},B_{H1},B_{H2})$ and the computational load is reduced to only $10!/7!/3!+7!/4!/3!+4!/2!/2!=161$ comparisons.\footnote{If  the Higgs buckets are filled before the top buckets, the combination is further
reduced to $10!/8!/2!+8!/6!/2!+6!/3!/3!=93$. However, it would increase the wrong assignments for both the signal and background. See also discussion in Section~\ref{subsec:metric}.}
In reality the number of jets will often be larger than the minimum of ten, 
where the speedup indicated here becomes even more prominent.
A detailed description of our specific algorithm is given in the next section.
We address possible issues related to the explicit decoupling of the buckets in Section~\ref{subsec:metric}.

\section{Benchmark scenario: Ten-jet final state from vectorlike top pair production}
\label{sec:analysis}
\subsection{Model and event generation}
\label{sec:analysis:model}

In this section, we illustrate our approach by using a simple model with a vectorlike top partner.
We extend the SM Lagrangian by adding a vectorlike top $T$ that interacts with the SM top $t$ and Higgs $H$,\footnote{In general, there can also be a model-dependent term $\lambda H \bar{t}\gamma_5 T + h.c. $ in the Lagrangian. Here we assume $\lambda=0$ for simplicity.}
\begin{equation}
\mathcal{L}=\mathcal{L}_{\rm SM} + \bar{T}(i\slashed{D}-m_{T})T + y_{T}H\bar{t}T + h.c.~.
\end{equation} 
We assume that the vectorlike top decays exclusively to a top and a Higgs. The mass of the vectorlike top in consideration is $m_T=0.8-1.2$ TeV. The mass of the top is taken to be 173 GeV. The SM Higgs has mass 126 GeV and decays to $b\bar{b}$ with branching ratio 56 \%. 

\madgraphamc~2.2.1~\cite{1106.0522} is used for generating parton-level events, which undergo hadronization and showering through \pythia~6.426~\cite{hep-ph/0603175}. \delphes~3.1.2~\cite{1307.6346} with parameters tuned to the ATLAS detector is used for fast detector simulation.

The relevant SM background processes for our analysis (cf.~Sec.~\ref{sec:analysis:analysis}) and their respective NLO K-factors are
$pp\to t\bar{t}$ (1.61~\cite{Kidonakis:2008mu}), $pp\to t\bar{t}b\bar{b}$ (1.77~\cite{Bevilacqua:2009zn}), $pp\to t\bar{t} H$ (1.10~\cite{Frederix:2011zi}), and $pp\to bb\bar{b}\bar{b}$ (1.40~\cite{Worek:2013zwa}).
All final-state top quarks and Higgs bosons are decayed hadronically within \madgraphamc.
The following generator-level cuts are imposed:
minimum transverse momentum of each outgoing parton $p_\perp\geq 20\GeV$,
angular separation between outgoing light quarks and between a light quark and a bottom quark $\Delta R_{jj},\Delta R_{jb}\geq 0.2$,
and angular separation between a pair of bottom quarks $\Delta R_{bb}\geq 0.4$.
The latter cut is imposed to guarantee sufficient $b$ separation to employ statistically independent $b$ quark tagging.
The overall scalar transverse momentum is imposed $H_T^\text{parton level}\geq 1\TeV$, consistent with a similar (but stronger) cut at analysis level, cf.~Eq.~(\ref{eq:ht_cut}).\footnote{To determine the respective cross-section at large scalar transverse momentum, we cut on events generated with $H_T^\text{parton level}\geq 500\GeV$ to achieve better accuracy.
Only for plotting we also  generate $t\bar{t}$ and $bb\bar{b}\bar{b}$ events with $H_T^\text{parton level}\geq 1.2\TeV$.}
The cut on $H_T^\text{parton level}$ guarantees a reasonably large fraction of events in the signal regions.
Note that this parton level cut on $H_T$ makes it difficult to generate events at NLO,
because it acts differently on processes with additional jets at matrix element level
(the set of partons which contribute to the sum is different).
Matching of matrix element with additional jets is also difficult for the same reason.
Therefore, we generate background events at LO without matching to higher multiplicities at matrix element level. Thus, the absolute numbers of the background events should be taken with a grain of salt.
The generated signal events do not suffer from this approximation.

\subsection{The analysis}
\label{sec:analysis:analysis}

We present an analysis that aims to identify the fully hadronic final state $t\bar{t}HH$ from vectorlike top pair production.
We do not rely on large-radius ``fat'' jets and their substructure, which has become 
a standard approach whenever boosted heavy particles are involved.
Conversely, the approach presented here focuses on separately resolved (small-radius) jets and is intended as a proof-of-concept in a realistic and relevant process.

The proposed analysis consists of the following steps, each of which is described in detail in the remainder of this subsection.
\begin{enumerate}
 \item Event preselection cuts:\\
 Scalar transverse momentum $H_T\geq 1400\GeV$ and number of $b$-tagged jets $\# b\geq 4$.
 \item Jet reconstruction and cut $\#\text{jets}\geq 10$. Here, we use several different benchmark algorithms including the mass-jump algorithm.
 \item Assignment of jets to the four buckets $B_{t1}$, $B_{t2}$, $B_{H1}$, and $B_{H2}$ and cuts.
 \item Kinematic reconstruction of the vectorlike tops, depending on the number of identified top and Higgs buckets.
\end{enumerate}

\subsubsection{Event preselection}

The decay cascade of a heavy vectorlike top pair leads to an energy deposit of $H_T\sim\mathcal{O}(2m_T)$ in the detector.
$H_T$ is the scalar transverse momentum, defined as
\begin{align}
 H_T\equiv\sum_{\text{jets } j} p_{\perp}^{(j)} \,.
\end{align}
We require
\begin{align}
 H_T\geq 1400\GeV
 \label{eq:ht_cut}
\end{align}
to retain the majority of signal events for a vectorlike top with $m_T\sim 1\TeV$ while strongly suppressing all non-resonant background processes.
In the event preselection, 
jets are clustered with the anti-$k_T$ algorithm as implemented in \fastjet~3.0.6~\cite{1111.6097} with parameters $R=0.4$ and $p_\perp\geq 20\GeV$.
Note that the jets are reconstructed differently after the preselection.

As the signal process contains six $b$ quarks in the final state, we also cut on the number of bottom tags.
Tagging is performed by \delphes using the jets defined above.
We select a conservative working point where 70\% of $b$-initiated jets are identified correctly, $\epsilon_\text{tag}=0.70$,
and assume the mistag rates of charm-initiated jets to be $\epsilon_\text{mis}^{(c)}=0.10$,
and $\epsilon_\text{mis}^{(udsg)}=0.01$ for light (quark- or gluon-initiated) jets.
Cutting on
\begin{align}
 \# b\geq 4
\end{align}
reduces the relevant backgrounds to $b$-rich processes with high multiplicity,
$pp\to t\bar{t}$, $pp\to t\bar{t}b\bar{b}$, $pp\to bb\bar{b}\bar{b}$, and $pp\to t\bar{t}H$.

\subsubsection{Jet reconstruction}

Jets are reconstructed from all calorimeter towers that lie within $|\eta|<4.9$.
To avoid ``chopped'' jets at the boundary of the detector, we require $|\eta_\text{jet}|<4.0$ so that all jets are sufficiently central.
The key ingredient to this analysis is the choice of jet clustering algorithm.
In our study, we adopt the following benchmark algorithms and compare them:

\begin{itemize}
\item A C/A-like mass-jump clustering algorithm with parameters
\begin{align}
\text{[MJ06] : } & (\, R=0.6 \,,\quad p_\perp\geq 25\GeV \,,\quad \theta=0.7 \,,\quad \mu=50\GeV \,) \,.
\label{eq:parameters_mj06}
 \\
\text{[MJ10] : } & (\, R=1.0 \,,\quad p_\perp\geq 25\GeV \,,\quad \theta=0.7 \,,\quad \mu=50\GeV \,) \,. \label{eq:parameters_mj10}
\end{align}
\item
A standard setup with the Cambridge-Aachen algorithm and commonly used clustering parameters,
\begin{align}
\text{[CA03] : } & (\, R=0.3  \,,\quad p_\perp\geq 25\GeV \,) \,. \label{eq:parameters_ca}
 \\
\text{[CA04] : } & (\, R=0.4 \,,\quad p_\perp\geq 25\GeV \,) \,.
\end{align}
\end{itemize}
The minimum jet $p_\perp$ is set to be the same to allow for easy comparison of the results.
The additional veto parameters specific to mass-jump clustering, $\theta$ and $\mu$ in Eqs.~(\ref{eq:parameters_mj06}) and (\ref{eq:parameters_mj10}), are motivated by the results obtained from boosted top quarks~\cite{Stoll:2014hsa}.
The mass-jump veto leads to jets whose effective radius can vary, and
it is this inherent flexibility that will lead to improved results compared to standard jet clustering with fixed angular size.

Because some jets reconstructed with the mass-jump algorithm may experience a very large effective radius,\footnote{This effect will be investigated later, cf.~Fig.~\ref{fig:last_dij}.} contamination from pile-up and underlying event can pose problems in a realistic environment.
We therefore apply a trimming~\cite{Krohn:2009th} stage.
For each jet $j$, the constituents are re-clustered with a smaller radius $R_\text{trim}$, yielding hard and possibly also soft subjets.
The jet is then re-built only from the subjets $i$ that are hard enough,
\begin{align}
 p_{\perp,i} > f_\text{trim}\ p_{\perp,j} \,.
\end{align}
We choose $R_\text{trim}=0.2$ and $f_\text{trim}=0.03$ as suggested in Ref.~\cite{Krohn:2009th}.
Trimming is applied to all the benchmark points, MJ06, MJ10, CA03, and CA04.

After the jets are reconstructed, we require
\begin{align}
 \# \text{jets} \geq 10 \,,
\end{align}
three for each top quark and two for each Higgs boson.

\subsubsection{Bucket construction and tagging}

In order to keep the combinatorial choices of this multi-jet process at a manageable level, we make use of the idea of buckets~\cite{1302.6238,1310.6034}. First of all, the first top bucket $B_{t1}$ is filled with the three jets that minimize
\begin{align}
 \Delta = |m_\text{bucket} - m_t| \,.
 \label{eq:bucket_metric}
\end{align}
Here and for all other buckets, we limit the allowed jet
combinations to those which fulfill
\begin{align}
 p_{\perp, \text{bucket}} \geq 200\GeV \,.
 \label{eq:bucket_pt}
\end{align}
This prevents wrong assignments from widely separated low-energy jets, which are possible due to the sheer number of possible choices.
In addition, only combinations with minimum mutual jet separation
\begin{align}
 \Delta R(j,j) \geq 0.3
 \label{eq:bucket_r}
\end{align}
are considered, because smaller distances cannot reasonably be resolved by the hadronic calorimeter any more.
This cut is also consistent with the cuts applied on generator level, cf.~Sec.~\ref{sec:analysis:model}.
Note that we do not impose an upper cut on angular spread of the top decay products, as is implicitly done in all substructure methods which rely on a fat jet of fixed radius.
Also note that Eq.~\eqref{eq:bucket_r} does not restrict the analysis if all jets are mutually separated by more than $\Delta R(j,j) = 0.3$, i.e.~the fixed-$R$ setups CA03 and CA04 are unaffected.
If two top subjets are very close-by and merge in the CA03 setup, even in the ideal case that the MJ algorithm can resolve them separately, they could not contribute to the same bucket.
In this sense the cut helps to allow a fair comparison between the mass-jump setups and the Cambridge-Aachen setups.

After the first top bucket has been fixed, out of the remaining jets the second top bucket $B_{t2}$ is filled with three jets, then the first Higgs bucket $B_{H1}$ with two jets, and finally $B_{H2}$ again with two jets.
This course of action corresponds to a global metric with explicitly decoupled buckets as defined in Eqs.~(\ref{eq:bucket_global_metric}) and~(\ref{eq:bucket_decoupled_metric}).
Again for each bucket, out of all possible jet combinations that fulfill Eqs.~(\ref{eq:bucket_pt}) and (\ref{eq:bucket_r}),
the combination with minimum metric (Eq.~(\ref{eq:bucket_metric}), where $m_t$ is replaced by $m_H$ for Higgs buckets) is selected.
Events where there is no viable jet combination for a bucket are negligibly rare.
Remaining jets are assigned to a fifth bucket $B_\text{ISR}$ and not further considered in our analysis.

Only after all buckets have been filled, cuts are applied.
For top candidate buckets, we require
\begin{gather}
 \Delta\leq 25\GeV \,, \label{eq:bucket_cut_delta}\\
 \left( \frac{m_W}{m_t}\right) _\text{bucket} \in \frac{m_W}{m_t} \pm 15\% \,, \label{eq:bucket_cut_massratio}\\
 \left( \frac{m_{23}}{m_{t}} \right) _\text{bucket} \geq 0.35 \,. \label{eq:bucket_cut_m23}
\end{gather}
Eq.~(\ref{eq:bucket_cut_delta}) is a simple cut on the reconstructed top mass.
The mass ratio in the left-hand side of Eq.~(\ref{eq:bucket_cut_massratio}) is constructed from the two jets which best reconstruct the $W$ boson mass ($m_{W,\text{bucket}}$) and the total jet mass of the bucket ($m_{t,\text{bucket}}$), as proposed when the bucket algorithm was introduced in Ref.~\cite{1302.6238}.
The final cut in Eq.~(\ref{eq:bucket_cut_m23}) was introduced in the HEPTopTagger~\cite{1006.2833}, where $m_{23}$ is the combined mass of the two sub-leading jets in the bucket (in terms of $p_\perp$).
In our study, it helps to suppress top candidates whose momentum is dominated by one very hard prong.

Higgs candidate buckets have to fulfill
\begin{align}
 \Delta \leq 20\GeV \,.
\end{align}
The 4-momentum of the successful top or Higgs candidate is given by the momentum sum of the jets inside the bucket.

\subsubsection{Signal regions and kinematic reconstruction}

\begin{table}[t]
 \begin{center}
  \begin{tabular}{|c||c|c|c|}\hline
   & \hspace{0.5cm}SR1\hspace{0.5cm} &\hspace{0.5cm} SR2\hspace{0.5cm} &\hspace{0.5cm} SR3\hspace{0.5cm} \\ \hline
   number of tagged $t$ & = 1 & = 2 & = 2\\ \hline
   number of tagged $H$ & = 2 & = 1 & = 2\\ \hline
  \end{tabular}
  \caption{The three signal regions.}
  \label{tab:sr}
 \end{center}
\end{table}

We define three signal regions depending on the number of tagged buckets, see Tab.~\ref{tab:sr}.
In addition to event rates, we kinematically reconstruct the vectorlike top from the momenta of a tagged top quark and a Higgs boson to assess its invariant mass
\begin{align}
 M(t,H)=\sqrt{(p_t+p_H)^2} \,.
\end{align}
In the case of a fully reconstructed event (SR3), we choose between the two possible pairings,
$\{(t_1,H_1),(t_2,H_2)\}$ and $\{(t_1,H_2),(t_2,H_1)\}$,
such that the mass difference of the two vectorlike tops is minimal,
\begin{align}
 \min\left[ |M(t_1,H_1)-M(t_2,H_2)|, |M(t_1,H_2)-M(t_2,H_1)| \right] \,.
 \label{eq:mass_minimax}
\end{align}

The majority of events, however, falls into signal regions 1 and 2, and we are left with three tagged and one untagged bucket.
As the untagged bucket also contains a significant energy deposit, its momentum can be used as an estimate for the fourth particle.
Again we apply Eq.~(\ref{eq:mass_minimax}) to determine the correct pairing.
Only the vectorlike top that is reconstructed from two tagged buckets is further considered.

\subsection{Results}
\label{sec:analysis:results}

The cut flow and expected event numbers at the LHC14 with 100$\invfb$ integrated luminosity are shown in Tab.~\ref{tab:result_nofevents} for two benchmark setups MJ10 [Eq.~(\ref{eq:parameters_mj10})] and CA03 [Eq.~(\ref{eq:parameters_ca})].
For both setups, the $T\bar{T}$ signal outnumbers the SM background up to $m_T=800-900\GeV$ in signal region the SR2, and up to $m_T=1.1\TeV$ in the signal regions SR1 and SR3.
Note that the absolute numbers should be taken with great care due to the simplified event generation setup, cf.~Sec.~\ref{sec:analysis:model}. A comparison of the relative significances between the employed algorithms is less affected by the uncertainties, though.

\begin{table}[thbp]
 \begin{center}
  \begin{tabular}{|c|c|c|c|c|c|c||c||c|c|c|c|} \hline
   \multicolumn{2}{|c|}{Process} & \multicolumn{5}{|c|}{$T\bar{T}$} &b.g. & $bb\bar{b}\bar{b}$ & $t\bar{t}$ & $t\bar{t}b\bar{b}$ & $t\bar{t}H$   \\ \hline
    \multicolumn{2}{|c|}{} & 800\GeV & 900\GeV & 1.0\TeV & 1.1\TeV & 1.2\TeV & &  &  &  &    \\\hline
%    Cross section[fb] & 3.75 & 1.52 & & & & --- & 2.20$\times10^6$ & 1.39$\times10^5$ & 494 & 25.5  \\ \hline
%%
   \multicolumn{12}{l}{number of events for 100$\invfb$} \\ \hline
   \multicolumn{2}{|c|}{$H_T\geq1.4\TeV$} & 507 & 306 & 167 & 86.7 & 43.6 &   25600 & 4130 & 20600 & 772 & 52.9  \\\hline
   \multicolumn{2}{|c|}{$\# b \geq 4$} & 356 & 217 & 118 & 60.8 & 30.6 &   1730 & 990 & 506 & 218 & 16.4 \\\hline \hline
   & $\# \text{jets} \geq 10$ & 306 & 185 & 101 & 52.7 & 26.8 &   518 & 166 & 201 & 141 & 9.5 \\
   \cline{2-12} \cline{2-12}
   MJ10 & SR1 & 14.9 & 10.4 & 6.4 & 3.8 & 1.9 &    3.3 & 0.5 & 1.1 & 1.5 & 0.1 \\
   \cline{2-12}
   & SR2 & 36.5 & 20.7 & 11.5 & 5.7 & 2.7 &    22.2 & 1.8 & 11.4 & 8.1 & 0.9 \\
   \cline{2-12} 
   & SR3 & 10.5 & 6.0 & 3.9 & 2.2 & 1.0 &    2.1 & 0.1 & 1.1 & 0.8 & 0.1 \\\hline
   \multicolumn{12}{l}{} \\ \hline
   & $\# \text{jets} \geq 10$ & 282 & 172 & 90.6 & 45.6 & 22.9 &   392 & 121 & 145 & 118 & 7.7 \\
   \cline{2-12} \cline{2-12}
   CA03 & SR1 & 8.4 & 8.0 & 4.7 & 2.4 & 1.2 & 2.4 & 0.4 & 1.0 & 0.9 & 0.1 \\
   \cline{2-12}
   & SR2 & 24.4 & 14.1 & 6.8 & 3.6 & 1.8 & 11.1 & 0.8 & 5.3 & 4.5 & 0.5 \\
   \cline{2-12} 
   & SR3 & 5.6 & 2.9 & 1.8 & 1.1 & 0.4 & 0.8 & 0.0 & 0.4 & 0.4 & 0.1 \\\hline
 \end{tabular}
 \caption{
 Expected event numbers for two benchmark setups, mass-jump clustering MJ10 [Eq.~(\ref{eq:parameters_mj10})] and standard Cambridge-Aachen clustering CA03 [Eq.~(\ref{eq:parameters_ca})].
 Numbers are given for an integrated luminosity of 100$\invfb$ at the LHC with $\sqrt{s}=14\TeV$. Results for the signal are shown separately for different values of the vectorlike top mass ranging from 800\GeV\ to 1.2\TeV. All relevant background processes as well as their sum (``b.g.'') are given in the right-hand columns. The three signal regions (SR) are defined in Table~\ref{tab:sr}.}
 \label{tab:result_nofevents}
 \end{center}
\end{table}

We observe that event numbers are largest in SR2 (2 tagged top quarks, 1 tagged Higgs boson).
This  is particularly pronounced for the various background processes.
It can be understood by the order in which the four buckets are filled:
The top buckets are filled first and reconstruct the truth partons very well, as will be investigated in Sec.~\ref{sec:tagging}.
If jets originating from a Higgs boson are wrongly assigned to a top bucket, it becomes unlikely to fill both Higgs buckets from the remaining jets with masses within the mass window.
This effect is larger for background processes, among which only a vanishing fraction contains actual Higgs bosons at parton level except the $t\bar{t}H$ background.
Thus the Higgs buckets are dominantly filled from the remaining unrelated jets.

As can be seen in Tab.~\ref{tab:result_nofevents}, in 
the conventional clustering setup CA03, event numbers are considerably smaller than those obtained with mass-jump clustering MJ10, both for the signal and SM backgrounds.
This is already observable at the $\# \text{jets} \geq 10$ cut stage, and the difference becomes even larger when events in the final signal regions are compared.
Due to the fixed jet radius of CA03, hard prongs that are separated by a distance smaller than $R=0.3$ merge, and it is easily understood that the number of hard jets is naturally smaller than 
the one obtained from a (reasonable) mass-jump setup.
As our implementation of the bucket algorithm explicitly requires resolved constituent jets, those merged jets fail to reconstruct their hard resonance, leading to a large drop in event numbers in all signal regions.

\begin{figure}[ht]
 \begin{center}
  \includegraphics[width=0.48\textwidth]{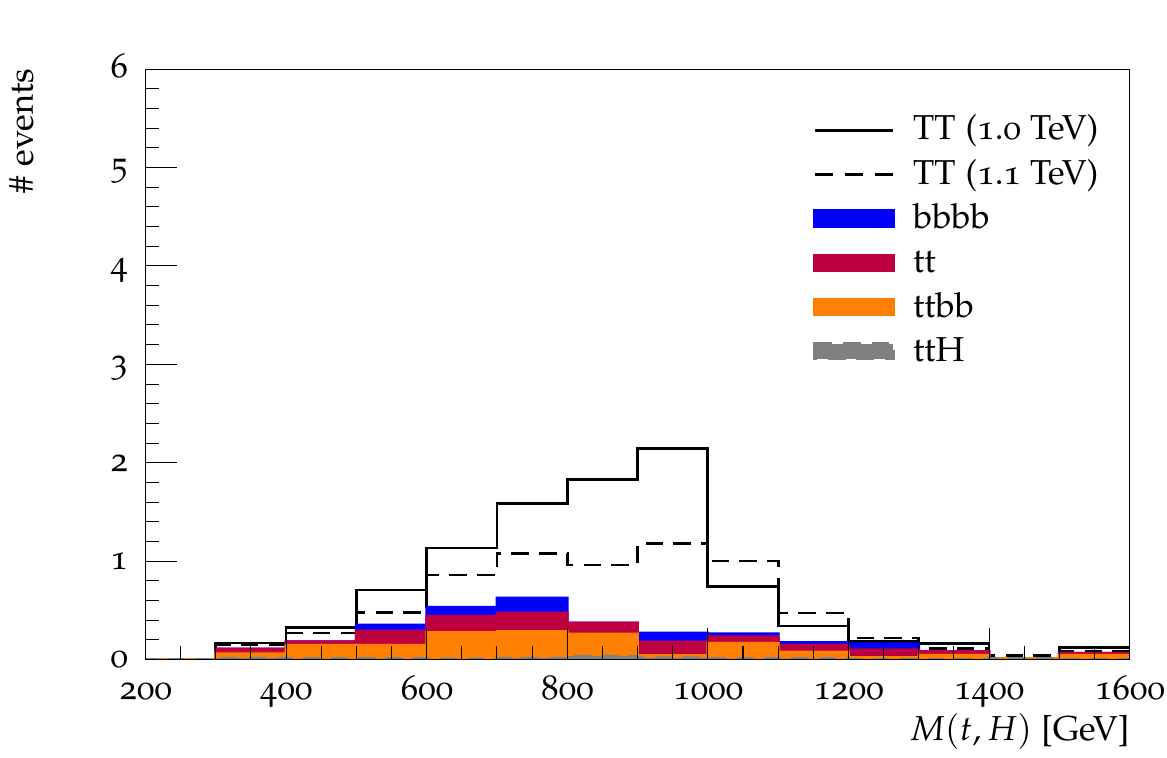}
  \includegraphics[width=0.48\textwidth]{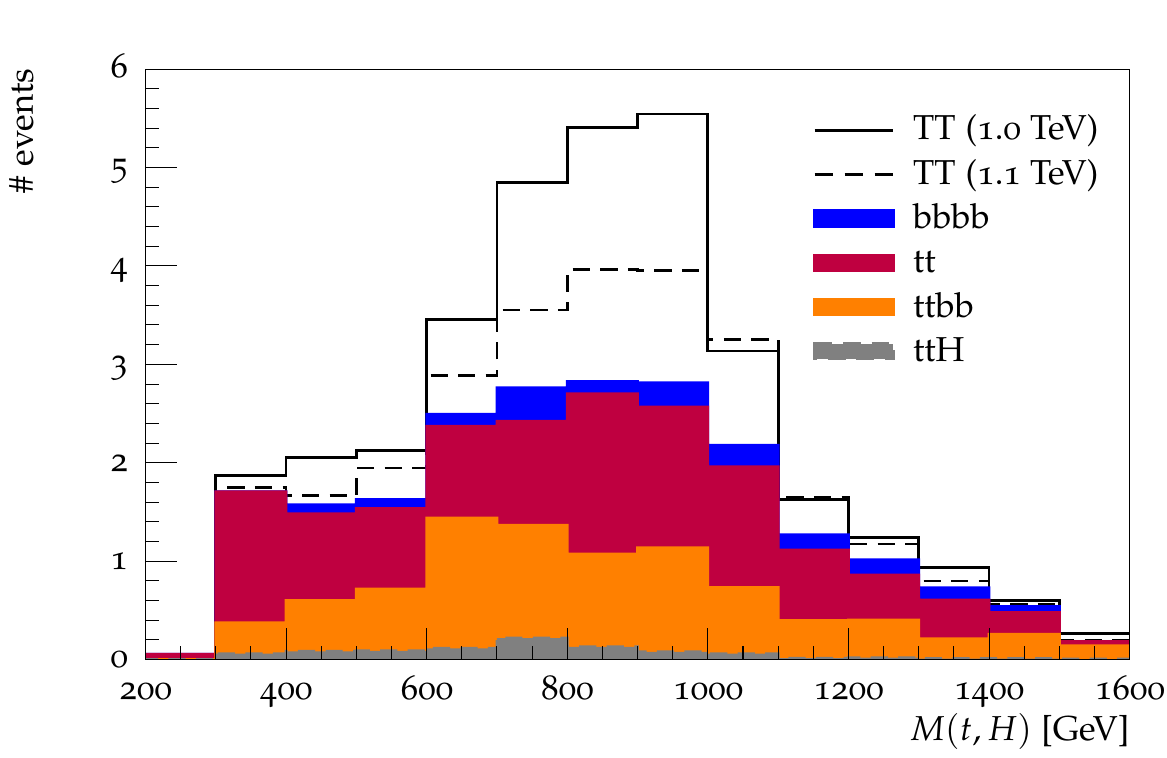}
  \includegraphics[width=0.48\textwidth]{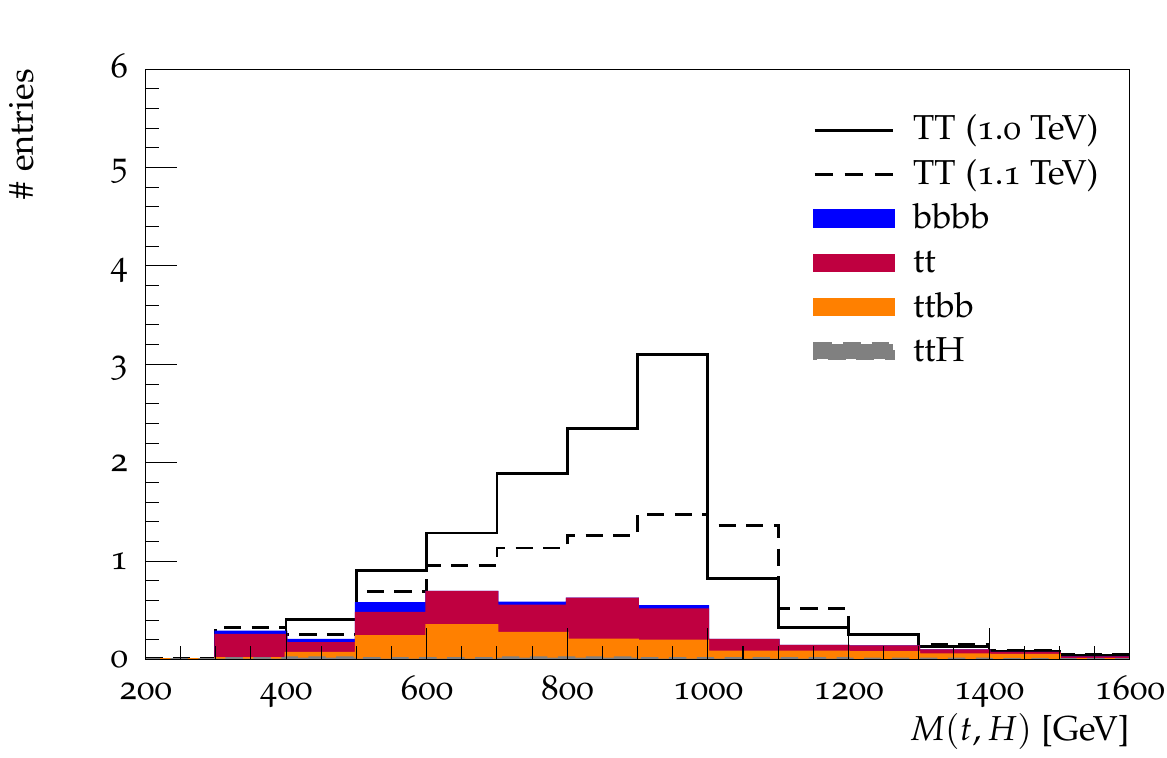}\\
 \end{center}
 \caption{Reconstructed mass of the vectorlike top $M(t,H)$ (for truth $m_T=1\TeV$ and $1.1\TeV$) with the MJ10 setup for SR1 (top left), SR2 (top right) and SR3 (bottom), for an integrated luminosity 100$\invfb$.
  The histograms are stacked.}
 \label{fig:reco_mass_mj10}
\end{figure}

In Fig.~\ref{fig:reco_mass_mj10}, we show the distributions of the vectorlike top mass, where stacked histograms of 
all three signal regions SR1 -- SR3 are presented.
(In SR3, each event gives two entries.)
The kinematic reconstruction of the vectorlike top works very well, as manifest in a clear peak in the figures.

In order to compare the different jet clustering setups, it is instructive to look at signal significance $S/\sqrt{B}$, which we take from the number of signal events $S$ and number of background events $B$ summed over all three signal regions.
Numbers are given in Tab.~\ref{tab:result_comparison} for all considered setups.
It is observed that among the standard clustering setups CA03 and CA04, the smaller jet radius yields better results.
The reason is that nearby prongs can only be separately resolved if the radius parameter is smaller than the mutual separation.
For mass-jump clustering MJ10 and MJ06, the opposite is true: the larger maximum jet radius gives more significant results.
Even in very busy final states some prongs end up fairly isolated, and they are more accurately reconstructed with larger jets.
Overall the mass-jump algorithm outperforms the fixed-radius conventional clustering.

\begin{figure}[t]
 \begin{center}
  \includegraphics[width=0.48\textwidth]{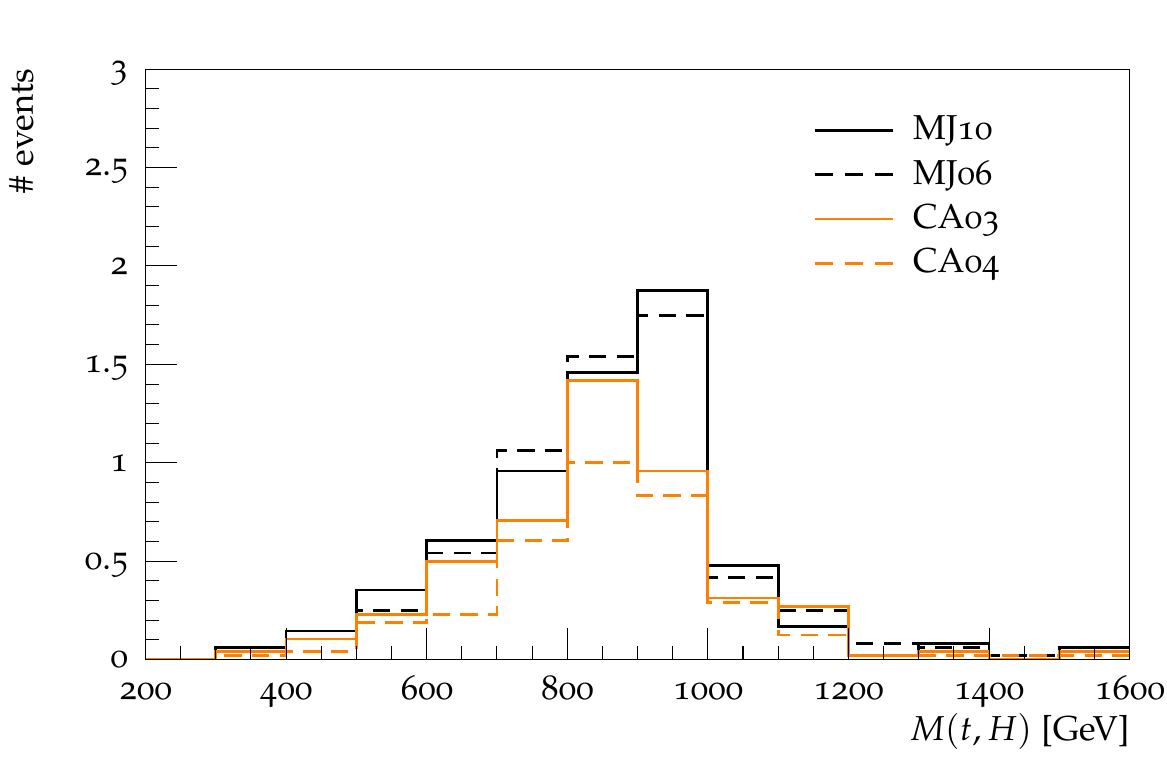}
  \includegraphics[width=0.48\textwidth]{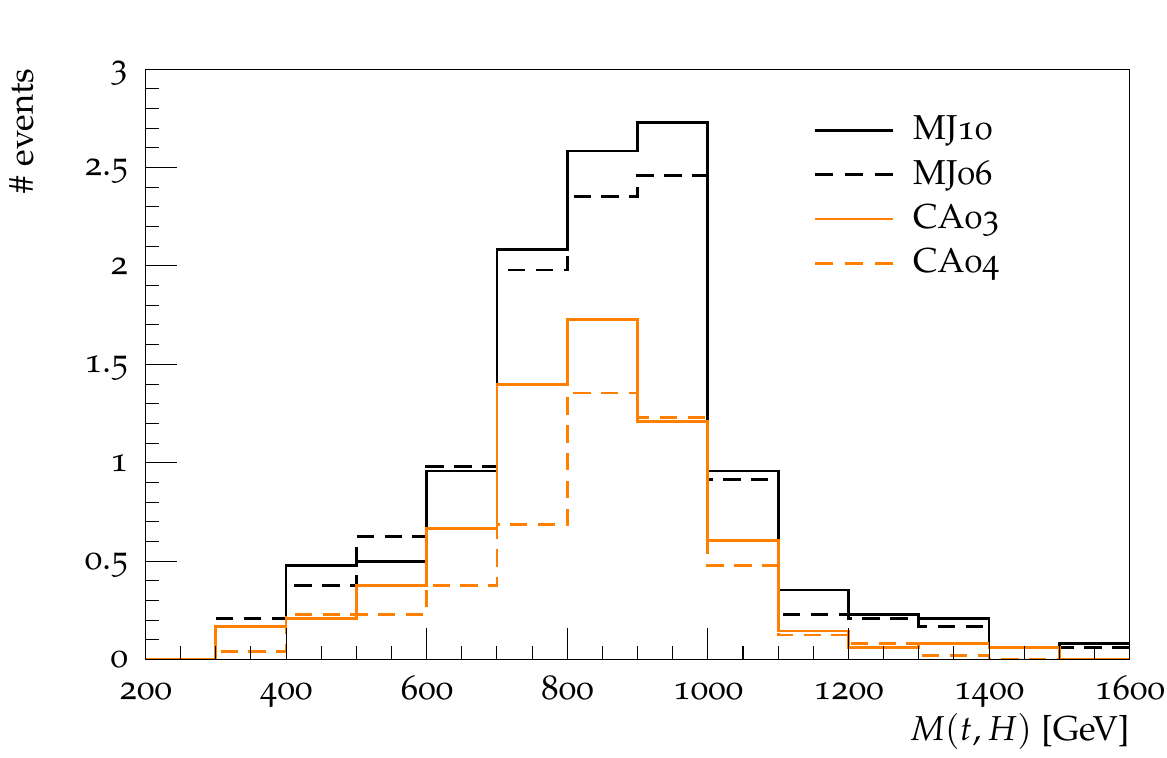}
  \includegraphics[width=0.48\textwidth]{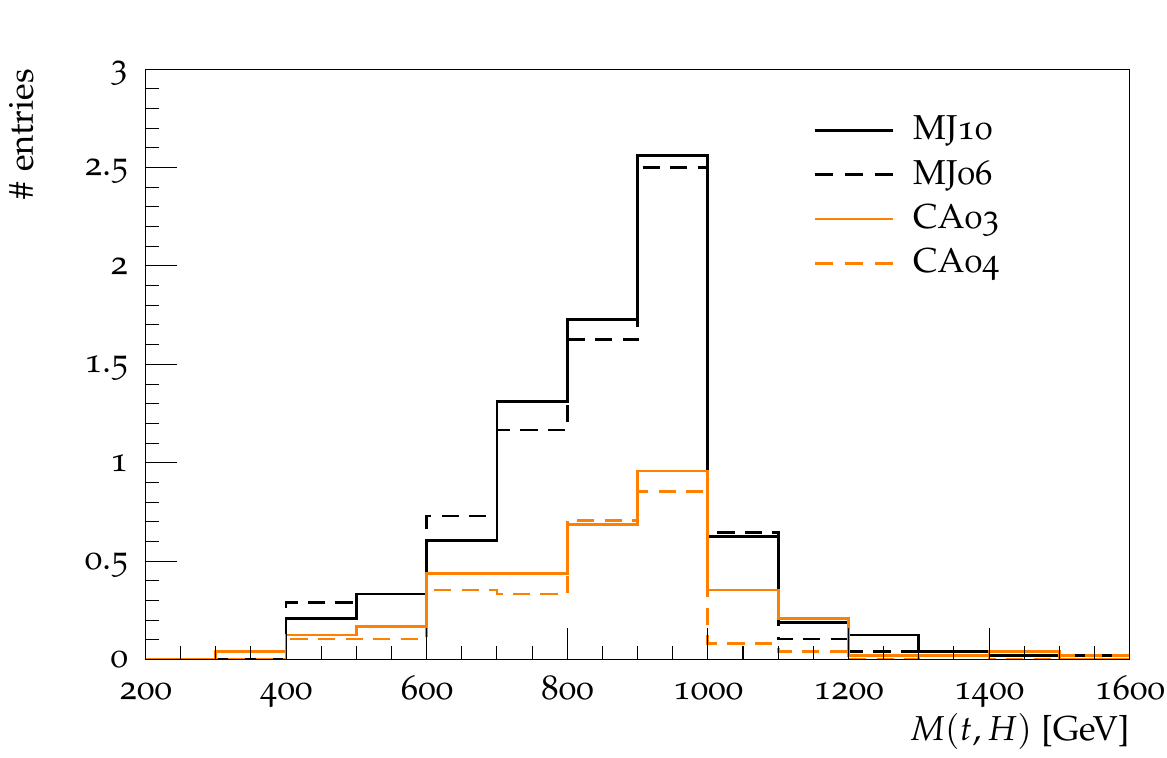}\\
 \end{center}
 \caption{Reconstructed $T$ mass (for truth $m_T=1\TeV$) in SR1 (top left), SR2 (top right) and SR3 (bottom) for different jet algorithms, 
 for an integrated luminosity $100\invfb$.
 }
 \label{fig:reco_mass_compare}
\end{figure}

The reconstructed signal mass (for truth $m_T=1\TeV$) is shown in Fig.~\ref{fig:reco_mass_compare} for all setups.
A peak is visible for all jet clustering setups, but for the fixed-radius algorithms CA03 and CA04 it is shifted to lower values in the SR1 and SR2.
The reconstruction is worst for the CA03 setup.
Only the analysis based on the mass-jump clustering can reproduce the mass of the heavy $T$ in all signal regions.
Independent of the specific clustering algorithm, the reconstructed mass peak has an edge around the true mass, with the majority of events experiencing a lower value.
This may be due to the fact that we do not explicitly include the leading gluon emission when the buckets are reconstructed.

Possible explanations for these observations and a comparison between standard Cambridge-Aachen and the mass-jump jets are given in the following subsection.

\begin{table}[t]
 \begin{center}
  \begin{tabular}{|c|c|c|c|c|c|} \hline
   $S/\sqrt{B}$ & 800\GeV & 900\GeV & 1.0\TeV & 1.1\TeV & 1.2\TeV \\\hline
   MJ10 & 11.77 & 7.05 & 4.13 & 2.22 & 1.07 \\\hline
   MJ06 & 11.38 & 6.96 & 4.06 & 2.16 & 1.02 \\\hline
   CA03 & 10.17 & 6.63 & 3.49 & 1.86 & 0.90 \\\hline
   CA04 & 11.06 & 5.91 & 3.36 & 1.51 & 0.61 \\\hline
  \end{tabular}
 \end{center}
 \caption{Comparison of significance $S/\sqrt{B}$ (number of signal events $S$ and number of background events $B$ summed over all three signal regions) for different jet algorithms and benchmark setups.}
 \label{tab:result_comparison}
\end{table}

\subsection{Comparison of jet clustering algorithms}
\label{sec:analysis:comparison}

\begin{figure}[ht]
 \begin{center}
  \includegraphics[width=0.45\textwidth]{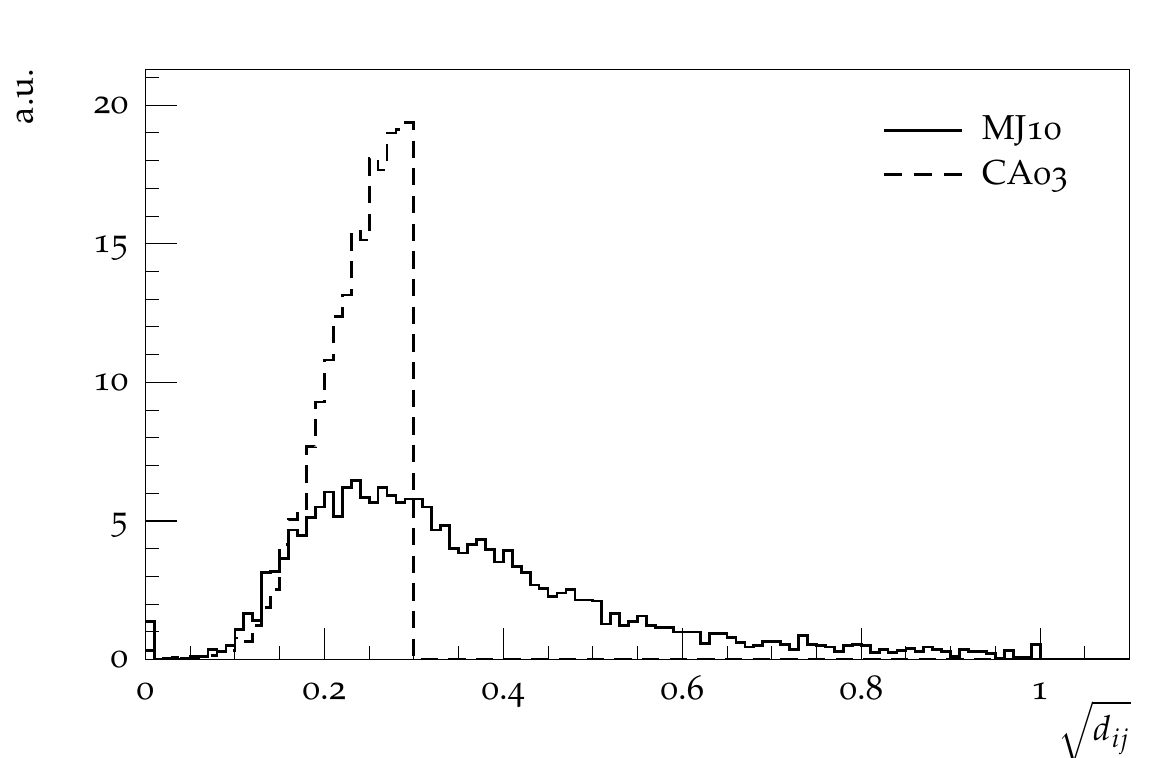}
  \includegraphics[width=0.45\textwidth]{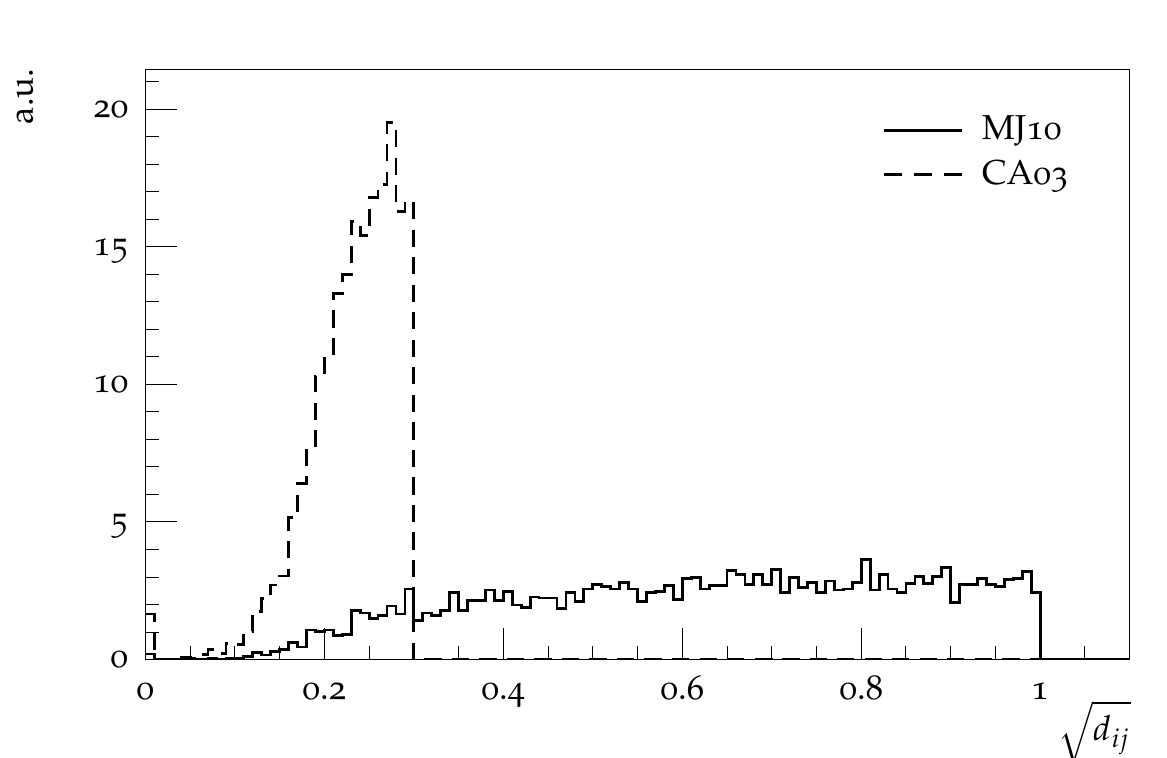}
 \end{center}
 \caption{$\Delta R=\sqrt{d_{ij}}$ of the last recombination step in the hardest (left) and tenth-hardest jet (right) in signal events with $m_T=1\TeV$ (arbitrary units).
 The solid lines depict values for jets clustered with the C/A-like mass-jump algorithm (MJ10), whereas jets clustered with the conventional C/A algorithm (CA03) are given by dashed lines.}
 \label{fig:last_dij}
\end{figure}

The results found in the previous subsection have mixed implications for  the ideal jet radius when standard fixed-$R$ clustering is employed.
CA03 yields larger overall significance than CA04, cf.~Tab.~\ref{tab:result_comparison}.
This is not surprising, as only a small radius can separately resolve hard prongs from boosted top and Higgs decays.
In terms of event numbers, this advantage seems to well compensate for possible splash-out, a loss of final-state radiation that falls outside the cone.
On the other hand, Fig.~\ref{fig:reco_mass_compare} shows that the reconstruction of the vectorlike top mass works better with a larger radius.
This shows the difficulty to find an optimal radius $R$ in the fixed-$R$ clustering algorithms.

Instead of employing a fixed clustering radius, the mass-jump algorithm was designed to separately resolve hard prongs at any distance scale if the terminating veto is called.
Fig.~\ref{fig:last_dij} shows the angular distance $\Delta R = \sqrt{d_{ij}}$ of the last recombination step in the hardest (left) and tenth-hardest jet (right).
Whereas for CA03 jets (dashed lines) the $\Delta R$ distribution peaks at the radius cut $R=0.3$ or slightly below, 
MJ10 jets (solid lines) observe much more variety.
For the hardest jet, $\Delta R$ has a peak at around 0.25 but can also take a large value, 
and for the tenth-hardest jet it has a broader and almost flat distribution.
%The $\Delta R$ distribution peaks around 0.25 for the hardest jet, and can also experience very large values as observed for the tenth-hardest jet.
This inherent flexibility constitutes the key to reconstructing the busy final state considered here.

The tail up to very large values in $\Delta R$ seen for the mass-jump tenth (soft) jets (Fig.~\ref{fig:last_dij} r.h.s.) 
may be a relic of the algorithm. In terms of significance, it was nevertheless observed that the overall performance is improved when such large radii are allowed.\footnote{This improvement is diminished due to trimming. By including a trimming stage, we assume that our results are not affected much if additional soft radiation from underlying event and pile-up are taken into account.
These effects should be included in a realistic study, but pile-up can only be reliably simulated by the experimental collaborations.
We assume that our results,  in particular the comparison between conventional jet clustering and mass-jump clustering, are still qualitatively valid in our simplified setup.}
These large-area jets can gather additional soft radiation, e.g.~soft gluon emissions, which can lead to a more accurate bucket mass. The fixed-$R$ setups CA03 and CA04 do not have these features, which could explain why the reconstructed $T$ mass in SR1 and SR2 is shifted to lower values for these algorithms. We speculate that a dedicated study of this effect may lead to improved taggers in this context, but it is beyond the scope of this paper.
\begin{figure}[ht]
 \begin{center}
  \includegraphics[width=0.45\textwidth]{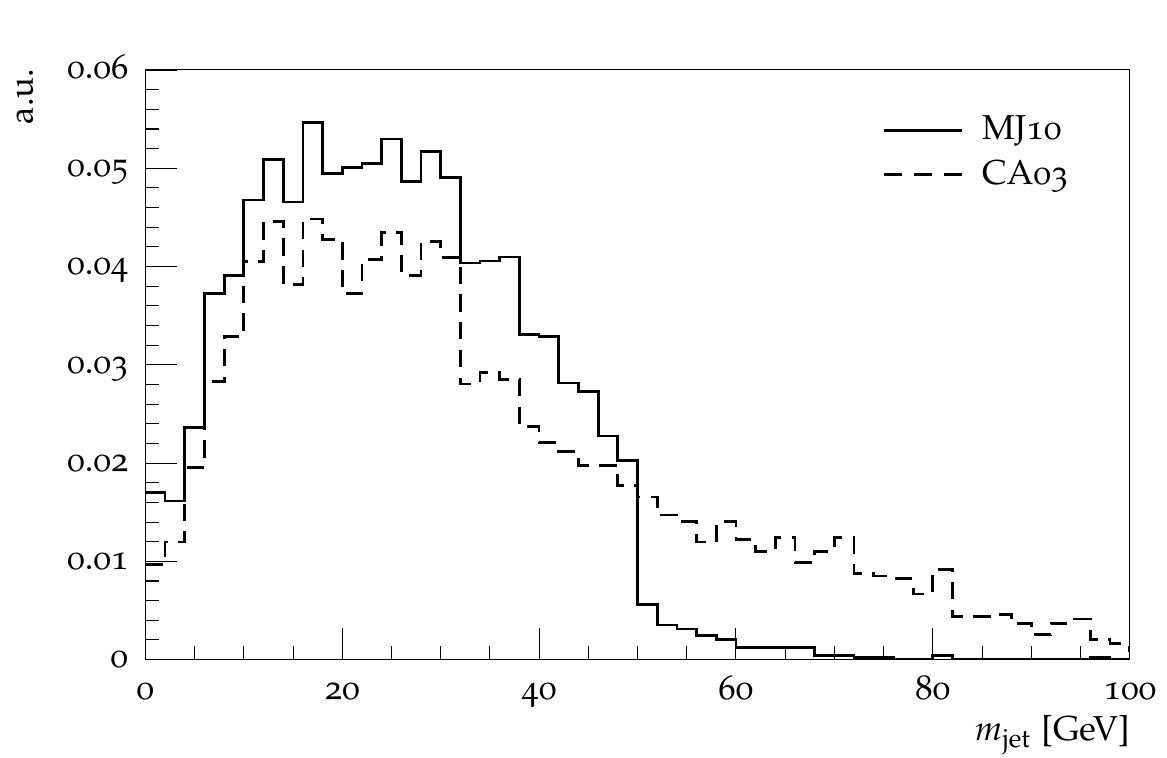}
  \includegraphics[width=0.45\textwidth]{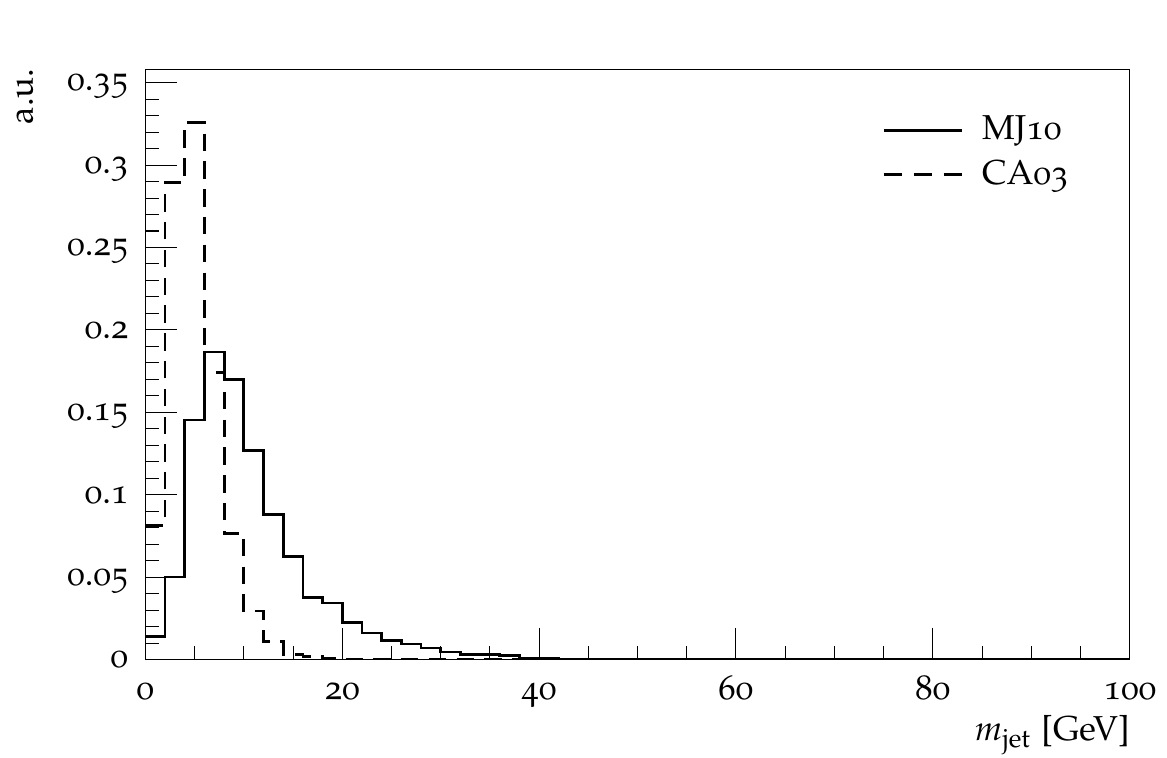}
 \end{center}
 \caption{Trimmed jet mass of the hardest (left) and tenth-hardest jet (right) in signal events with $m_T=1\TeV$ (arbitrary units).
 The solid lines depict values for jets clustered with the C/A-like mass-jump algorithm (MJ10), whereas jets clustered with the conventional C/A algorithm (CA03) are given by dashed lines.}
 \label{fig:jetmass}
\end{figure}

Fig.~\ref{fig:jetmass} shows the trimmed jet mass, again for the hardest (left) and tenth-hardest jet (right).
A fraction of events experiences a very heavy leading jet around $m_j=70\sim80\GeV$ in the CA03 setup, indicating that nearby hard prongs have merged.
The leading mass-jump jet, on the other hand, has a cutoff at $m_j=\mu=50\GeV$ due to the veto condition (cf.~Sec.~\ref{sec:recap:mj}), and very large jet masses are absent.
As plain jet mass roughly scales with $p_\perp \cdot R$, soft jets clustered with fixed-$R$ algorithms tend to be very light, as shown in the right panel of Fig.~\ref{fig:jetmass} for the CA03 setup.
However, final-state radiation of low-$p_\perp$ jets is less collimated
and ideally caught in jets with larger radius~\cite{Krohn:2009zg}.
In the same figure, it is seen that the tenth (soft) MJ10 jets are heavier
% and more physical, 
due to the larger effective jet radius as observed in Fig.~\ref{fig:last_dij}.

We conclude that the results found in Sec.~\ref{sec:analysis:results},
namely that a small jet radius can be of advantage for conventional fixed-$R$ clustering algorithms, whereas mass-jump clustering benefits from a very large maximum $R$,
can be explained by looking at jet merging scales and mass distributions.
For our process including four boosted resonances and a very busy final state, it is essential to find jets with a flexible algorithm.
The mass-jump algorithm avoids the problem of searching for a good compromise for the fixed jet radius parameter and leads to physically more appealing jets.
Consequently, it generally outperforms its standard fixed-$R$ counterpart, the Cambridge-Aachen algorithm, in the phenomenological analysis.

\subsection{On fat jet contamination}
\label{sec:analysis:fatjets}

In this subsection, we briefly discuss the expected performance of algorithms using fat jets in the present process, $pp\to T\bar{T}\to t\bar{t}HH\to 10\text{ jets}$.
In Sec.~\ref{sec:intro}, we argued that the fat jets are not well-separated
in such a busy hadronic final state, and  this problem is illustrated in Fig.~\ref{fig:contamination_deltar}
for the signal process with vectorlike top mass $m_T=1\TeV$.
The figure shows the angular distributions of the ten partonic (anti)quark final state (Monte Carlo truth) daughters.
The black solid (dashed) line shows the distribution of the largest angular distance between the truth daughters of one top quark (Higgs boson) found in an event.
The smallest distance between any truth daughters not coming from the same mother resonance is given by the red line.
It is observed that the distance between the nearest daughter particles coming from different mother resonances, 
$\Delta R(\text{NN})$, is typically smaller than the angular spread of a $t$ or $H$ decay, $\Delta R_{\text{max}}(t/H)$.
As a result, the fat jets will be contaminated.

\begin{figure}[t]
 \begin{center}
  \includegraphics[width=0.45\textwidth]{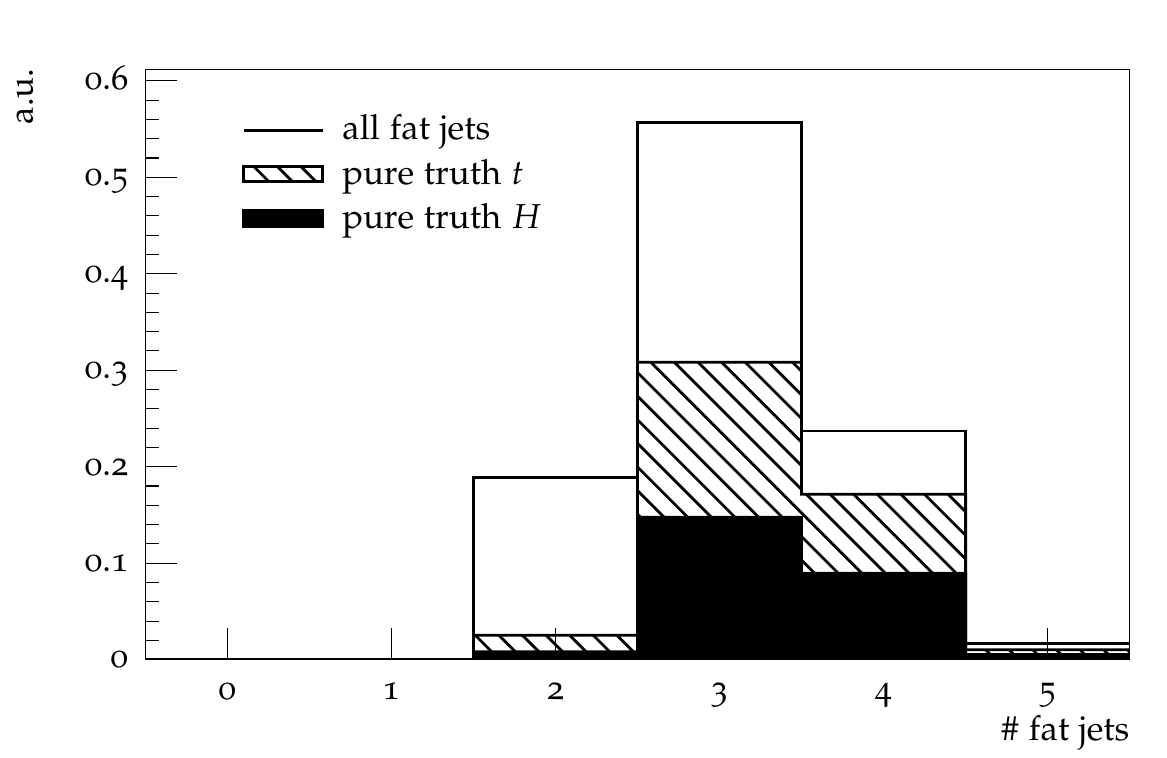}
  \includegraphics[width=0.45\textwidth]{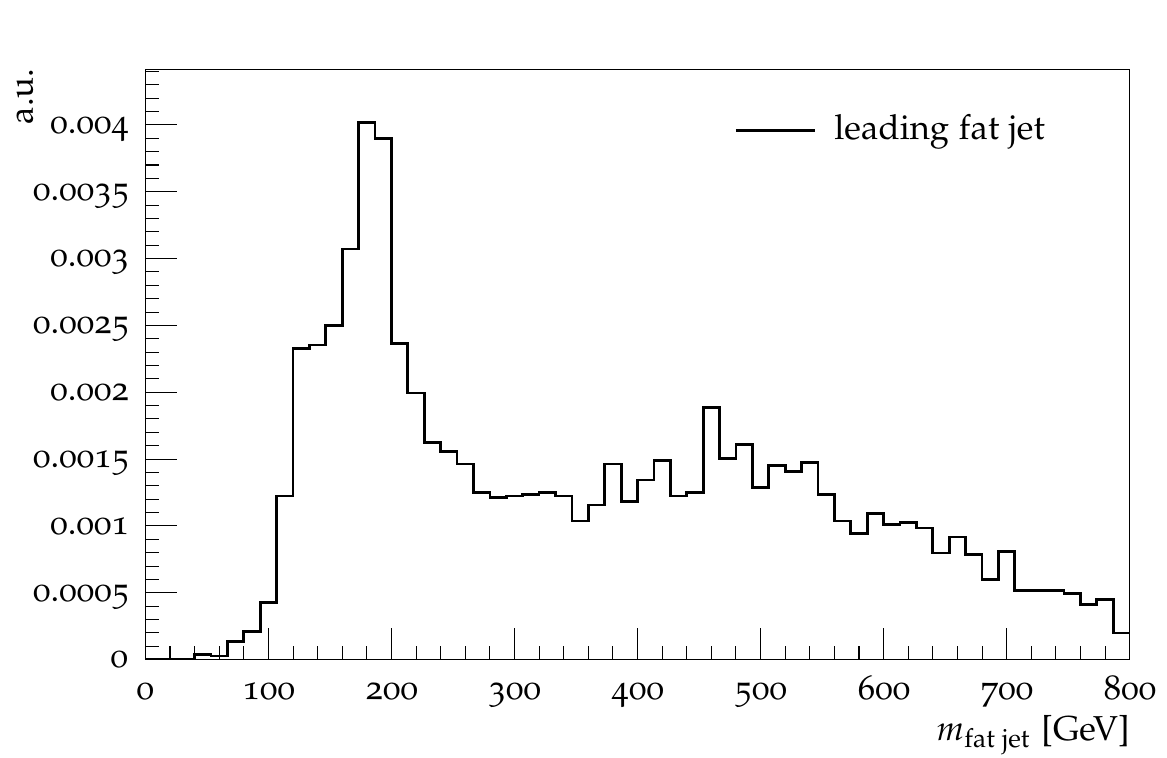}\\
  \includegraphics[width=0.45\textwidth]{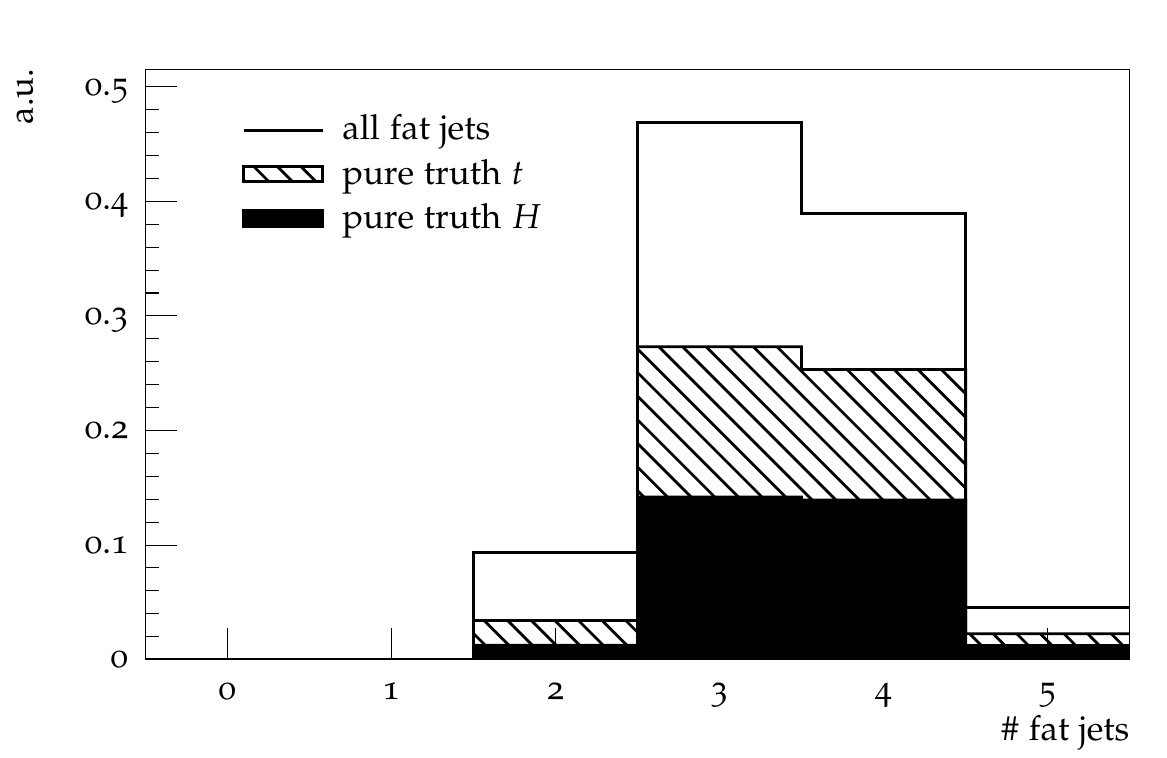}
  \includegraphics[width=0.45\textwidth]{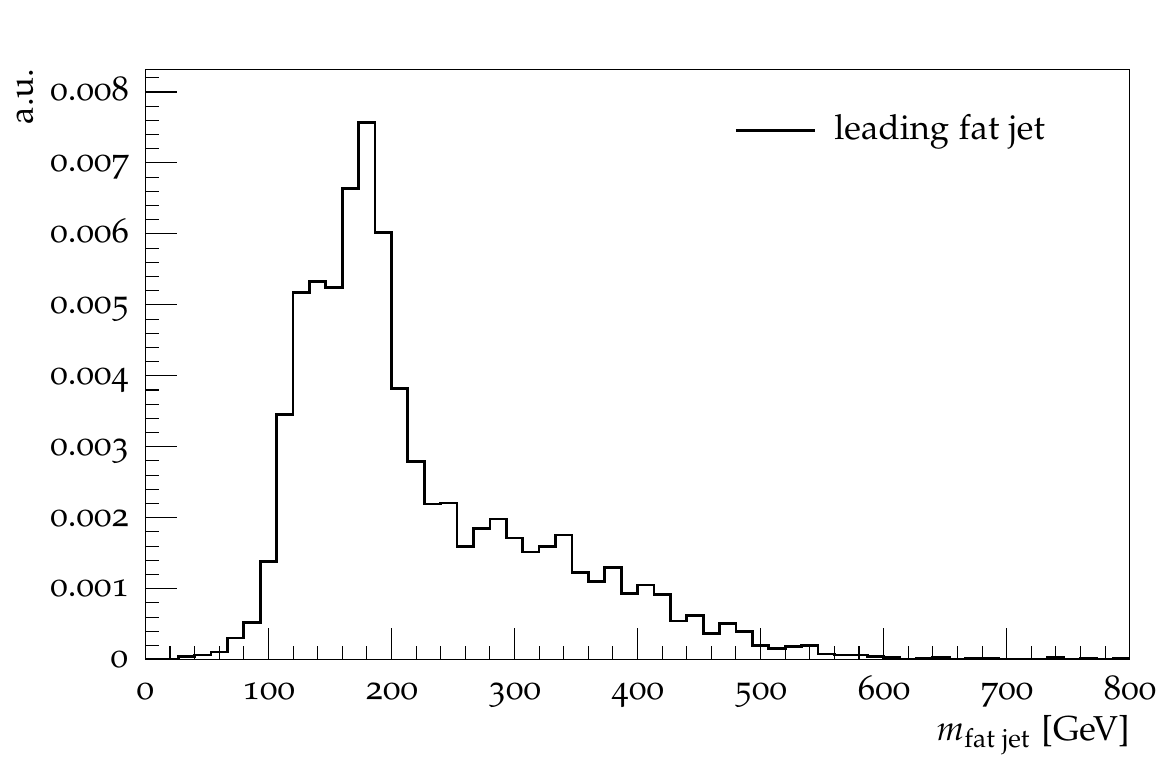}
 \end{center}
 \caption{The upper row shows the results for CA15 fat jets: The number of fat jets is given in the left plot.
 A fat jet is labelled pure truth $t$ ($H$) if all truth daughter partons of one top quark (Higgs boson) and no other truth daughter partons are ghost-associated.
%  within the fat jet. 
 For events in which three fat jets were found, the distribution of the plain jet mass of the leading fat jet is shown in the right plot.
 Lower row: The same plots for AKT10 fat jets.}
 \label{fig:contamination_fatjet}
\end{figure}

To be more specific, we take the default fat jet clustering parameters of the widely used HEPTopTagger~\cite{1006.2833}
\begin{align}
 \text{Cambridge--Aachen}: \qquad
 R^\text{fat jet} = 1.5 \qquad\text{and}\qquad
 p_\perp^\text{fat jet} \geq 200\GeV
\end{align}
and give some concrete results for the process
$pp\to T\bar{T}\to t\bar{t}HH\to 10~\text{jets}$ ($m_T=1\TeV$)
in Fig.~\ref{fig:contamination_fatjet} (CA15, upper panels).
A different choice of parameters is suggested 
in comparisons between boosted top tagging algorithms~\cite{Altheimer:2012mn},
\begin{align}
 \text{anti-}k_T: \qquad
 R^\text{fat jet} = 1.0 \qquad\text{and}\qquad
 p_\perp^\text{fat jet} \geq 200\GeV \,,
\end{align}
and we also give the  plots for these fat jets in Fig.~\ref{fig:contamination_fatjet} (AKT10, lower panel).

The upper panels imply that the clustering radius of the CA15 jets is too large in this situation.
The upper left panel shows the distribution of the number of fat jets.
In more than 50\% of events, only three fat jets are found, and even less in another 20\% (Fig.~\ref{fig:contamination_fatjet} upper left).
A fat jet is labelled ``pure truth" $t$ ($H$) if all truth daughter partons of one top quark (or Higgs boson) and no other truth daughter partons are ghost-associated~\cite{Cacciari:2008gn}.\footnote{The truth partons' momenta are rescaled to infinitesimal $p_\perp$ and energy while $\eta$ and $\phi$ are kept fixed (''ghosts``), and participate in jet clustering. Those partons that end up as constituents of a certain jet are called ghost-associated. Due to the vanishing energy of the ghosts, the final jets are unaffected.}
% within the fat jet. 
In each bin, the fraction of pure truth $t$ ($H$) fat jets are represented by the hatched (black) area. One can see that only a relatively small fraction of the fat jets are pure truth ones.
% While events with four fat jets are rare (~25\%), of these fat jets more than 70\% are pure.
A phenomenological study would have to rely on events with only three fat jets, and the plain jet mass of those leading fat jets is depicted in the upper right panel.
There is a large tail towards very large jet masses, which suggests that there is a significant amount of splash-in from jets not coming from the same $t$ or $H$ resonance.

The second setup (AKT10) with a smaller-radius fat jets, 
shown in the lower row of Fig.~\ref{fig:contamination_fatjet},
 behaves better in this respect.
In roughly 40\% of events the correct number of four fat jets is identified (lower left), although three-jet events are still dominant.
For the events with three fat jets, the leading jet mass is shown in the lower left panel.
It can be seen that the distribution still shows a tail, but now large jet masses are much less present than in CA15 jets, implying fewer contamination through splash-in.
On the other hand, when we compare the fraction of pure jets in events with four fat jets between the two setups, we observe that AKT10 jets behave worse:
While almost 75\% of the respective CA15 fat jets are pure, this number is degraded to 65\% for AKT10.

We conclude that the study of this process is difficult if we rely on fat jets.\footnote{For $m_T\lesssim 900\GeV$, an analysis based on fat jets can still reconstruct the vectorlike top~\cite{Endo:2014bsa}. 
An experimental analysis of the same process also relies on fat jets~\cite{Khachatryan:2015axa}.}
The problem of insufficient separation of the boosted resonances ($t$ and $H$) is not generically avoided 
even if a different fat jet radius is chosen -- the smallest distance between truth daughters from different mothers is typically smaller than the angular spread of the top quark and Higgs boson, as shown in Fig.~\ref{fig:contamination_deltar}.
There is no apparent solution to this contamination within fat jet algorithms, and the choice of clustering algorithm and parameters is related to finding a balance between splash-in (the fat jet contains energy deposit from a different resonance, too) and splash-out (the fat jet does not contain all radiation from a given resonance).
As demonstrated in the main part of this paper, 
this problem is reduced when the mass-jump algorithm is used.

\section{Performance of top and Higgs tagging}
\label{sec:tagging}

In this section, we investigate the performance of top and Higgs tagging in our approach
with the mass-jump (and the Cambridge-Aachen) clustering algorithms. We also briefly comment on the metric of the decoupled buckets 
in Sec.~\ref{subsec:metric}.

It should be emphasized that the
tagging efficiencies and the quality of reconstruction strongly depend on the considered physical processes as well as the event generator by which the samples have been produced.
This is even more true in our study, as the performance of top/Higgs tagging is affected by the hadronic activity from other top and Higgs decay products in the candidate's vicinity.
Tagging efficiencies and reconstruction qualities of the present canonical tagging algorithms for boosted resonances are usually evaluated for isolated fat jets
(see e.g.~Refs.~\cite{Altheimer:2012mn,Adams:2015hiv}), 
which reduces the dependence on the specific process and makes it possible to compare the results between different algorithms.
This condition is not satisfied in our benchmark analysis and therefore the results for top and Higgs tagging can hardly be related to other algorithms.
In addition, the strong weighting of the global buckets metric in Eq.~(\ref{eq:bucket_global_metric}) naturally leads to the first top bucket being much better reconstructed than the second.
Similarly, the reconstruction quality of the top quarks is generally better than that of the Higgs bosons.

Despite those precautional warnings, the results presented here can serve as a benchmark for other processes with a similarly busy final state.

\subsection{Reconstruction quality}

\begin{figure}[ht]
 \begin{center}
  \begin{minipage}{0.1\textwidth}
  \ 
  \end{minipage}
  \begin{minipage}{0.38\textwidth}
   top buckets
  \end{minipage}
  \begin{minipage}{0.48\textwidth}
   Higgs buckets
  \end{minipage}

  \begin{minipage}[t]{0.02\textwidth}
  \begin{sideways}\hspace{2cm}$m$\end{sideways}
  \end{minipage}
  \includegraphics[width=0.45\textwidth]{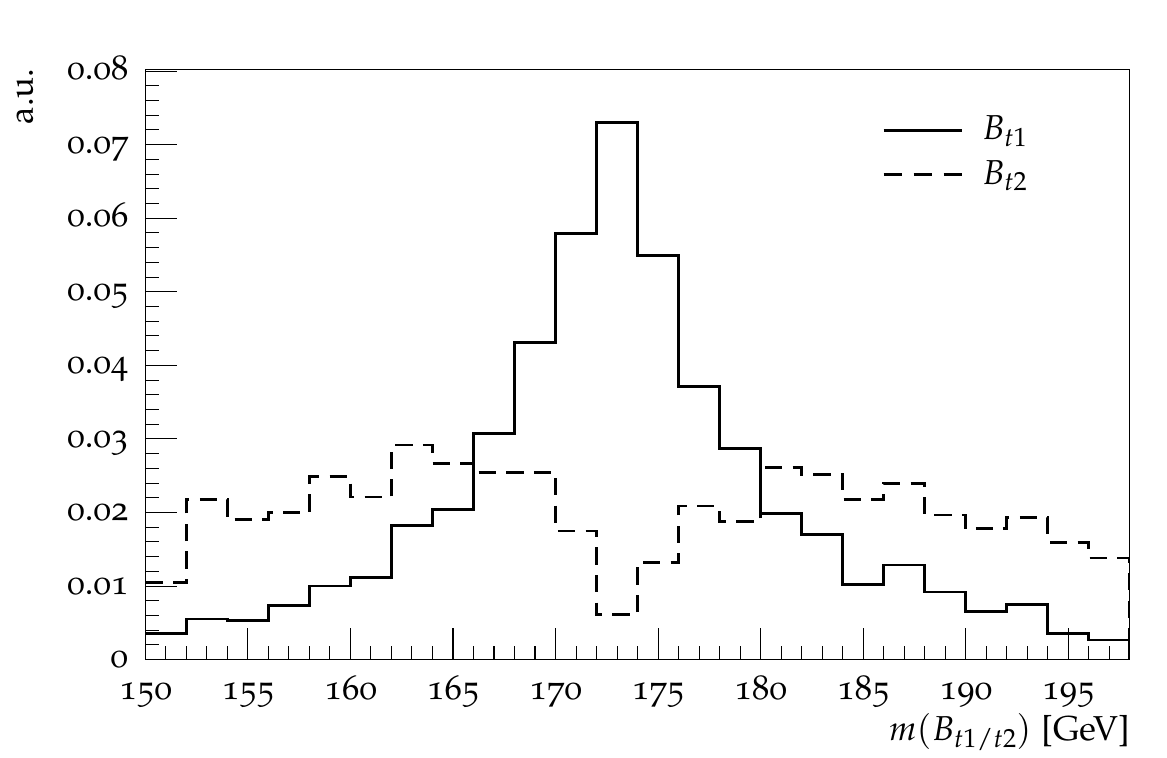}
  \includegraphics[width=0.45\textwidth]{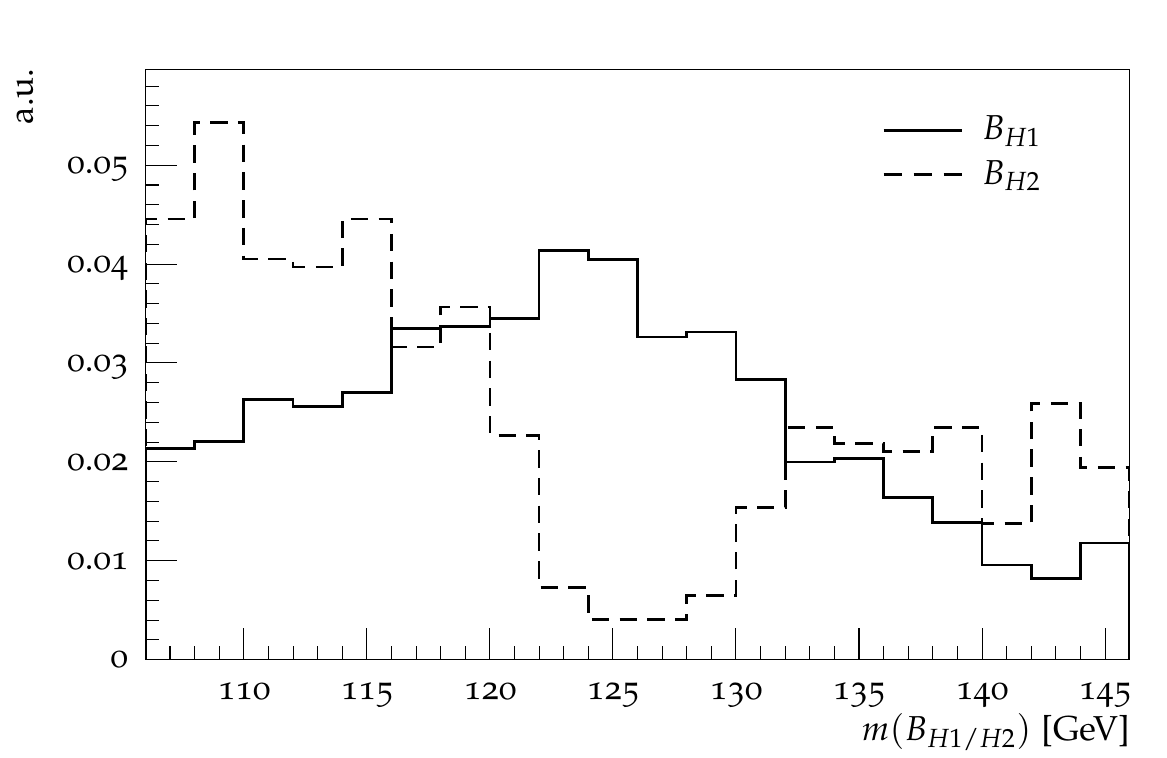}\\
  
  \begin{minipage}[t]{0.02\textwidth}
  \begin{sideways}\hspace{1.5cm}$\Delta p_\perp / p_\perp^\text{reco}$\end{sideways}
  \end{minipage}
  \includegraphics[width=0.45\textwidth]{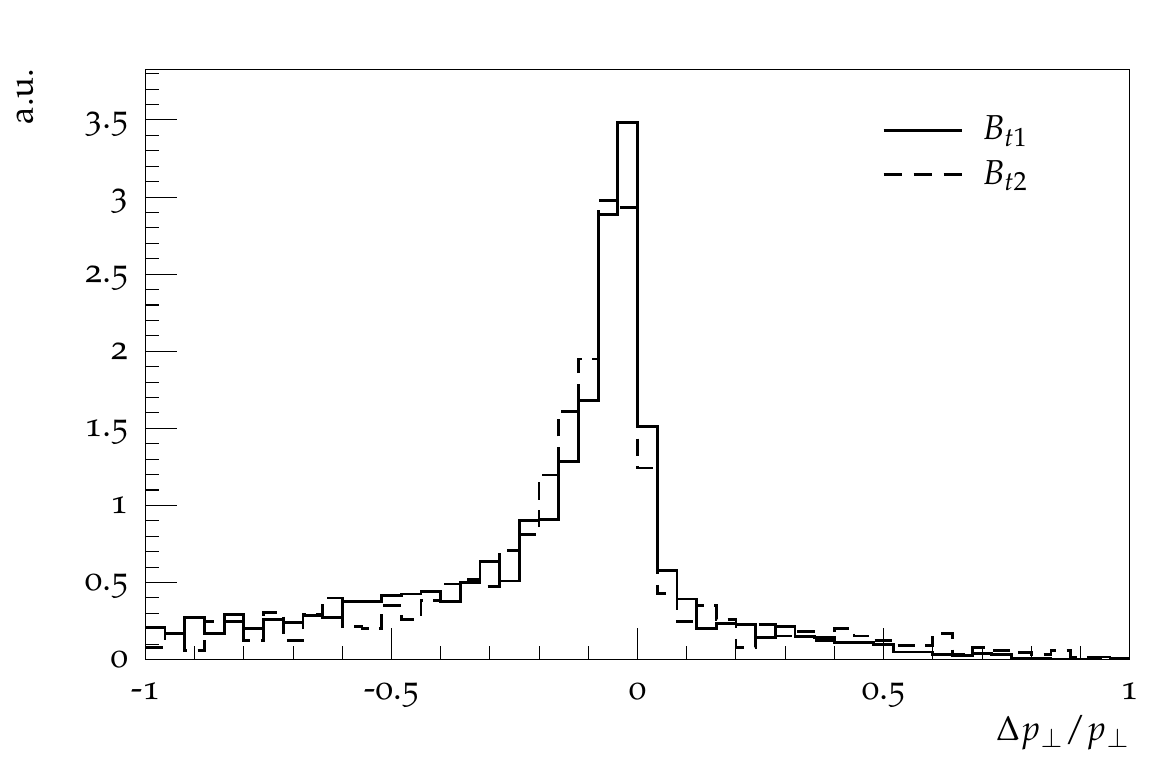}
  \includegraphics[width=0.45\textwidth]{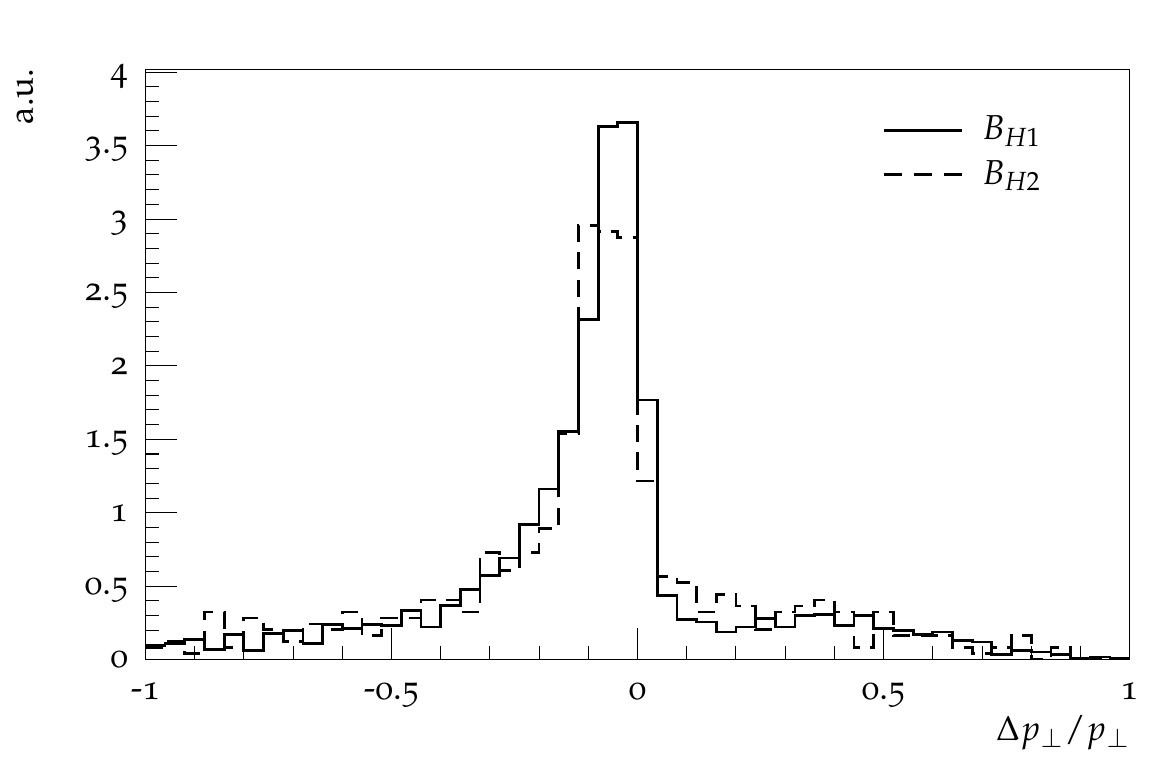}\\
  
  \begin{minipage}[t]{0.02\textwidth}
  \begin{sideways}\hspace{1cm}$\Delta R_{\text{reco},\text{parton}}$\end{sideways}
  \end{minipage}
  \includegraphics[width=0.45\textwidth]{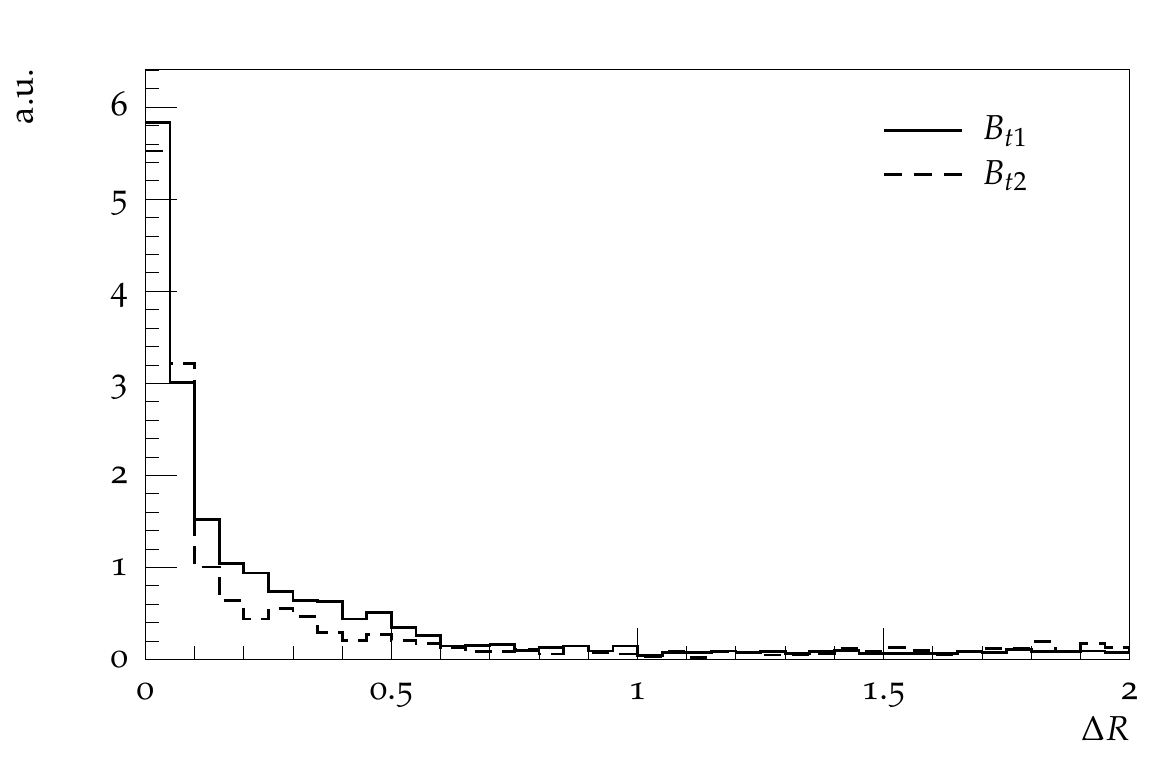}
  \includegraphics[width=0.45\textwidth]{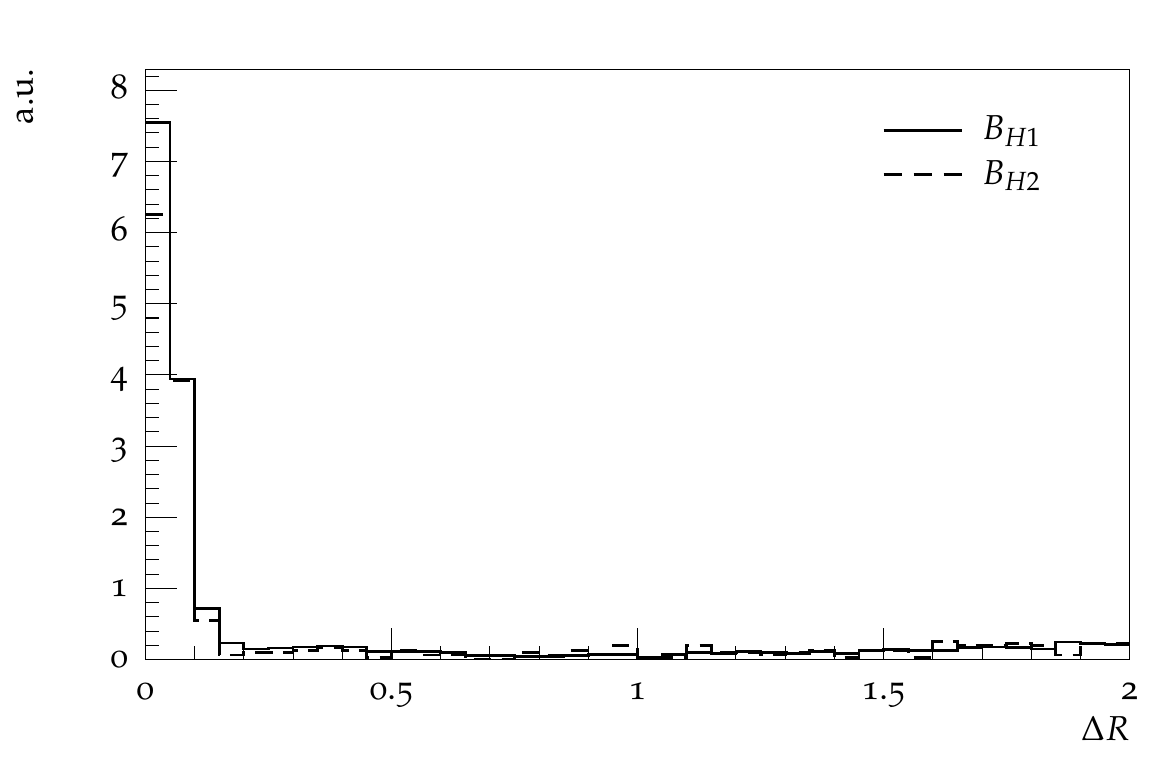}
 \end{center}
 \caption{Reconstruction quality of tagged top (left) and Higgs buckets (right) for the MJ10 setup,
 in the benchmark process  $pp\to T\bar{T}\to t\bar{t}HH\to 10\text{ jets}$
  at the LHC with $\sqrt{s}=14\TeV$ (arbitrary units).
 From top to bottom, the reconstructed mass, relative deviation in transverse momentum $\Delta p_\perp / p_\perp^\text{reco}$, and the angular distance $\Delta R_{\text{reco},\text{parton}}$ are shown.
 The solid curves show results for the first bucket, the dashed curves for the second bucket.
 }
 \label{fig:reco_quality_mj10}
\end{figure}

In Fig.~\ref{fig:reco_quality_mj10} we assess the quality of momentum reconstruction of the tagged buckets for the preferred MJ10 setup,
in the benchmark process  $pp\to T\bar{T}\to t\bar{t}HH\to 10\text{ jets}$
  at the LHC with $\sqrt{s}=14\TeV$.
The reconstructed masses of the top quarks and Higgs bosons are shown in the upper row.
Due to the ordering in the metric, the first bucket always gives a better reconstructed mass, leading to the dip for the second bucket.
Both for top and Higgs candidates, there is a central peak at the true mass value for the first buckets.
The top mass peak is much narrower, which is not surprising since Higgs buckets are filled by the remaining jets only after the two top buckets have been filled.
The middle and lower rows of Fig.~\ref{fig:reco_quality_mj10} show the deviation between the bucket momenta and the MC truth parton momenta in terms of two variables,
\begin{align}
 \frac{\Delta p_\perp}{p_\perp^\text{reco}} \equiv & \frac{p_\perp^\text{reco}-p_\perp^\text{parton}}{p_\perp^\text{reco}} \qquad \text{and}\\
 &\Delta R_{\text{reco},\text{parton}} \,.
\end{align}
A narrow peak around 0 is observed for both observables, for top as well as Higgs candidates.
In fact, most of the tagged buckets are reconstructed within 20\% in $\Delta p_\perp/p_\perp^\text{reco}$, and $\Delta R \leq 0.1$, with no significant differences between the respective first and second buckets.
We conclude that (i) the tagged buckets are built from the correct jets, and that (ii) these jets reconstruct the truth partons' momenta very well.

\begin{figure}[ht]
 \begin{center}
  \begin{minipage}{0.1\textwidth}
  \ 
  \end{minipage}
  \begin{minipage}{0.38\textwidth}
   top buckets
  \end{minipage}
  \begin{minipage}{0.48\textwidth}
   Higgs buckets
  \end{minipage}

  \begin{minipage}[t]{0.02\textwidth}
  \begin{sideways}\hspace{2cm}$m$\end{sideways}
  \end{minipage}
  \includegraphics[width=0.45\textwidth]{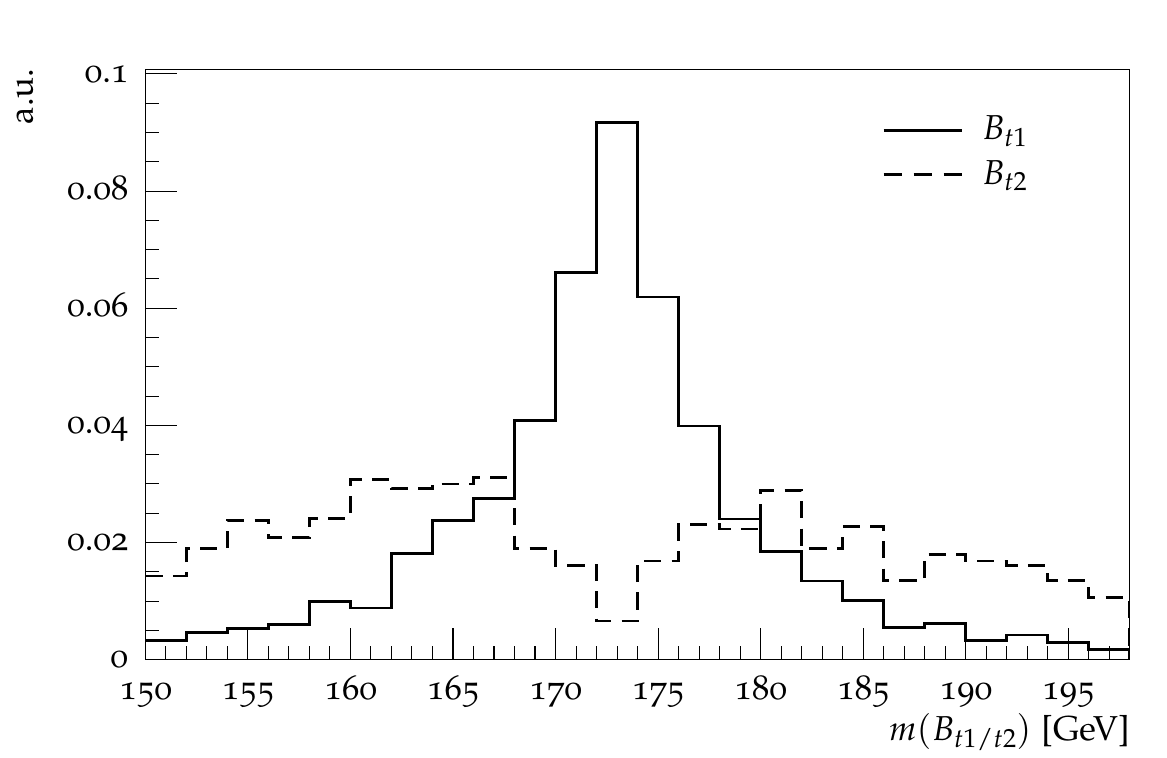}
  \includegraphics[width=0.45\textwidth]{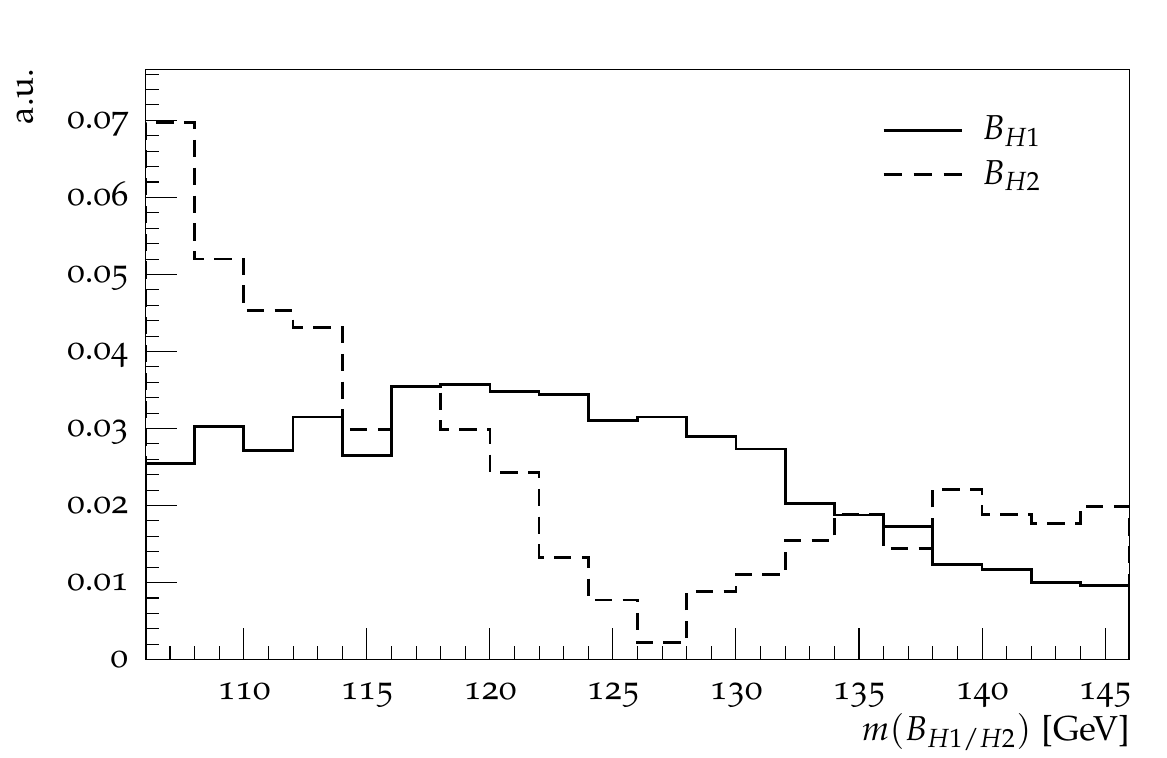}\\
  
  \begin{minipage}[t]{0.02\textwidth}
  \begin{sideways}\hspace{1.5cm}$\Delta p_\perp / p_\perp^\text{reco}$\end{sideways}
  \end{minipage}
  \includegraphics[width=0.45\textwidth]{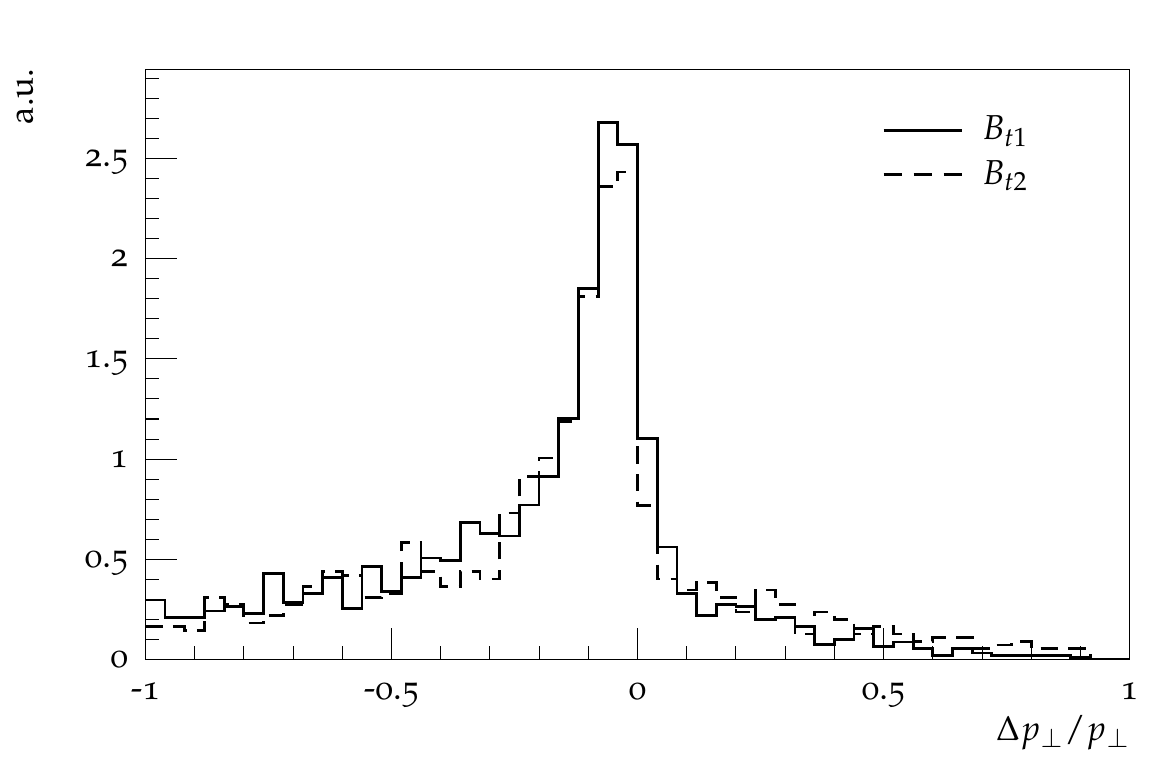}
  \includegraphics[width=0.45\textwidth]{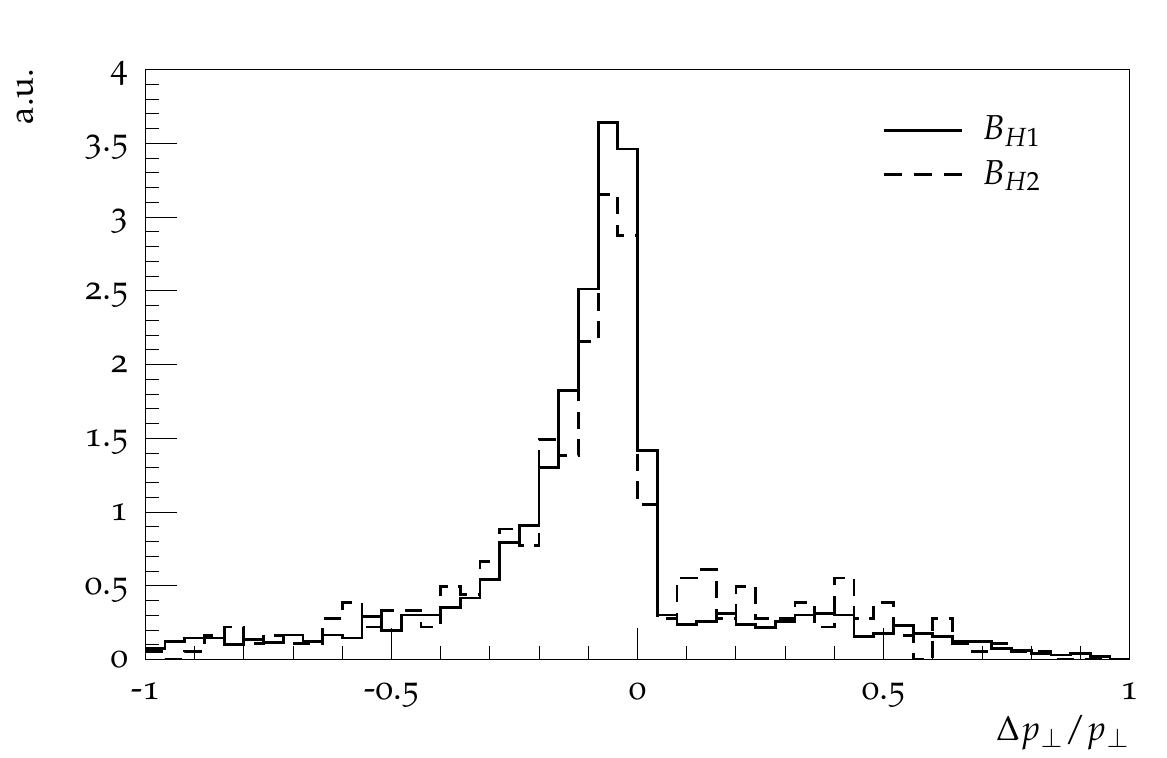}\\
  
  \begin{minipage}[t]{0.02\textwidth}
  \begin{sideways}\hspace{1cm}$\Delta R_{\text{reco},\text{parton}}$\end{sideways}
  \end{minipage}
  \includegraphics[width=0.45\textwidth]{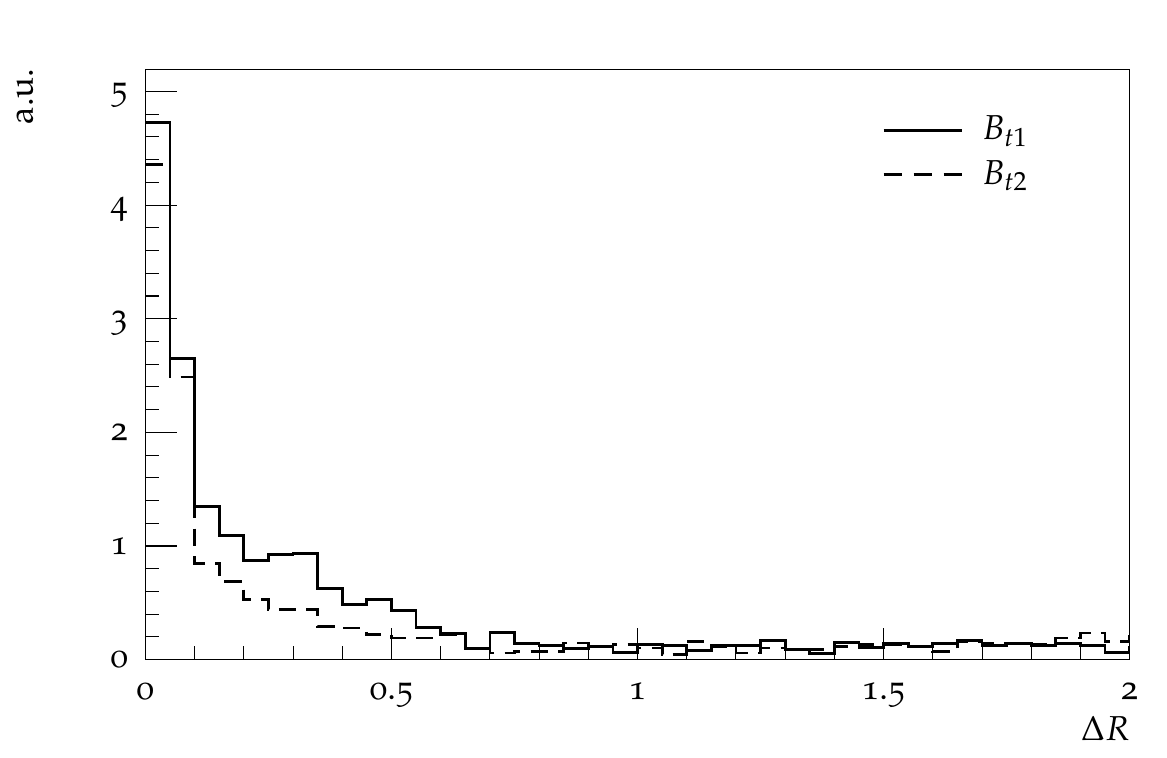}
  \includegraphics[width=0.45\textwidth]{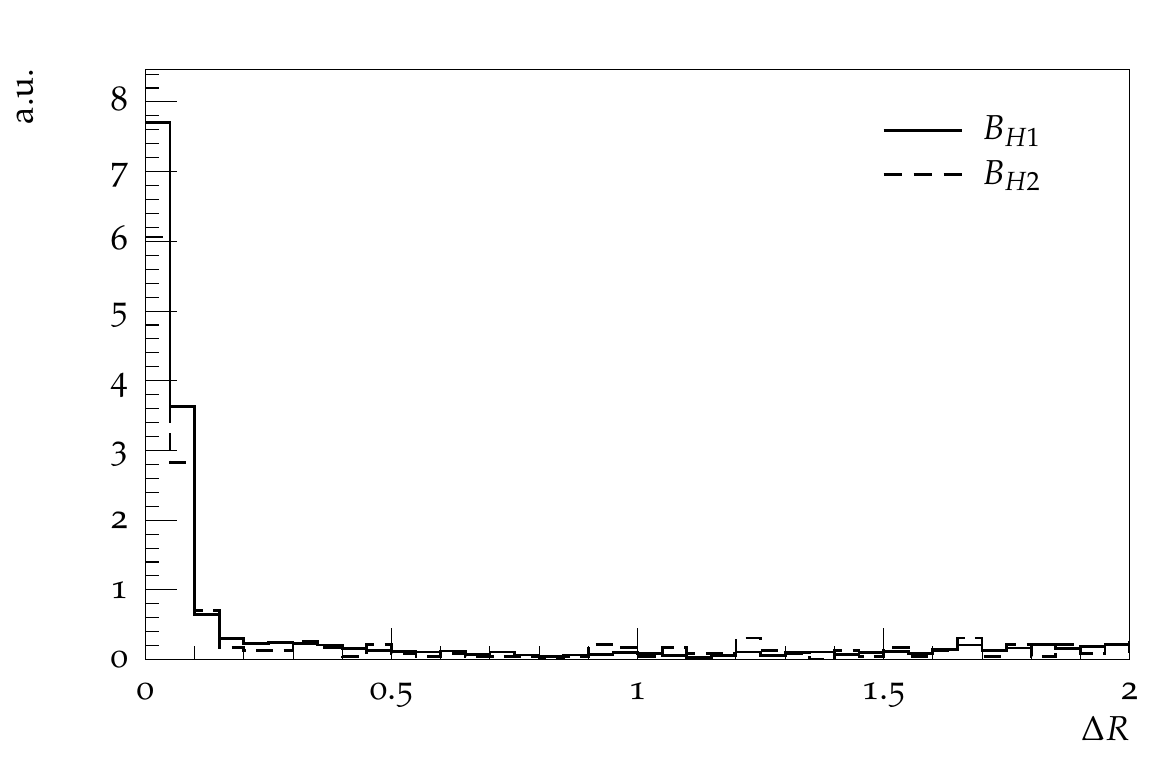}
 \end{center}
 \caption{The same as Fig.~\ref{fig:reco_quality_mj10} but for the CA03 setup.}
 \label{fig:reco_quality_ca03}
\end{figure}

For completeness, we also show the same results for the standard clustering benchmark setup CA03 in Fig.~\ref{fig:reco_quality_ca03}.
Note that, although the distributions look similar to the MJ10 setup, the total number of tagged buckets is significantly smaller.
The reconstructed Higgs bosons tend to have a broader peak, shifted to lower values in the CA03 setup.
As a result, the mass of the reconstructed vectorlike top is also shifted to lower values, as has been observed in Fig.~\ref{fig:reco_mass_compare}.
This is another indication that the mass-jump algorithm is better suited to this analysis, in addition to larger event numbers discussed in Sec.~\ref{sec:analysis:results}.

\begin{figure}
 \begin{center}
 \includegraphics[width=0.32\textwidth]{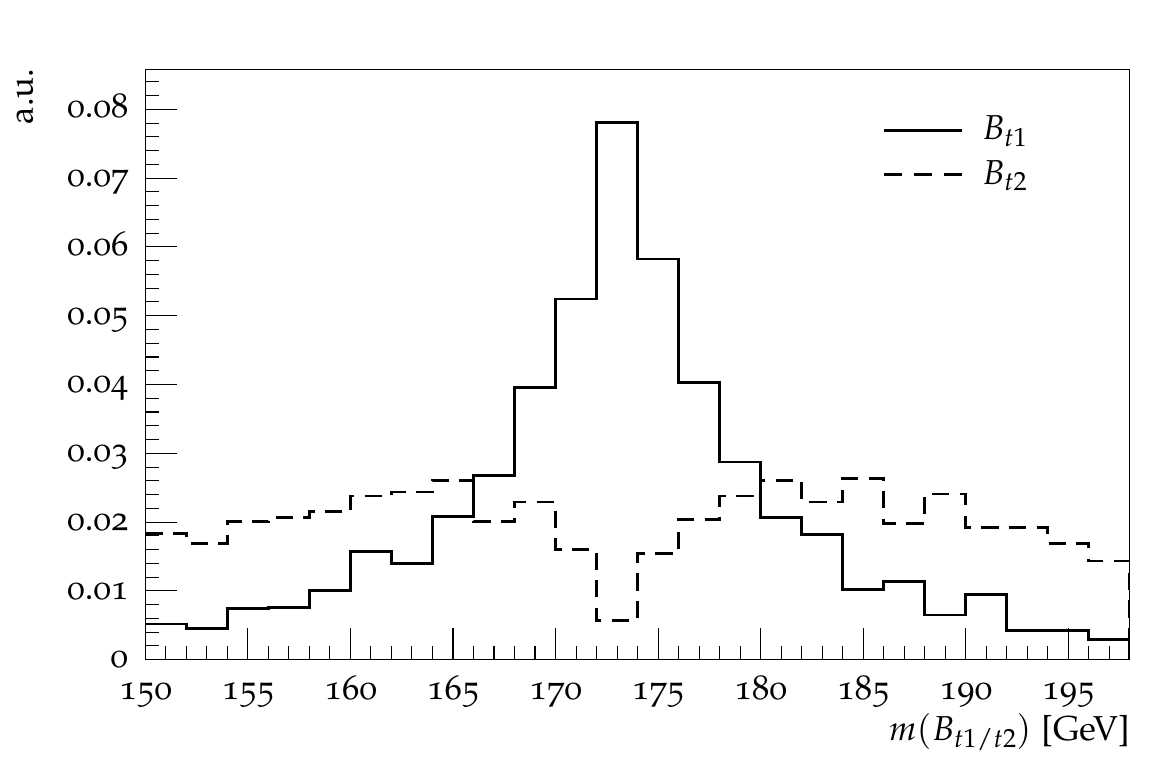}
 \includegraphics[width=0.32\textwidth]{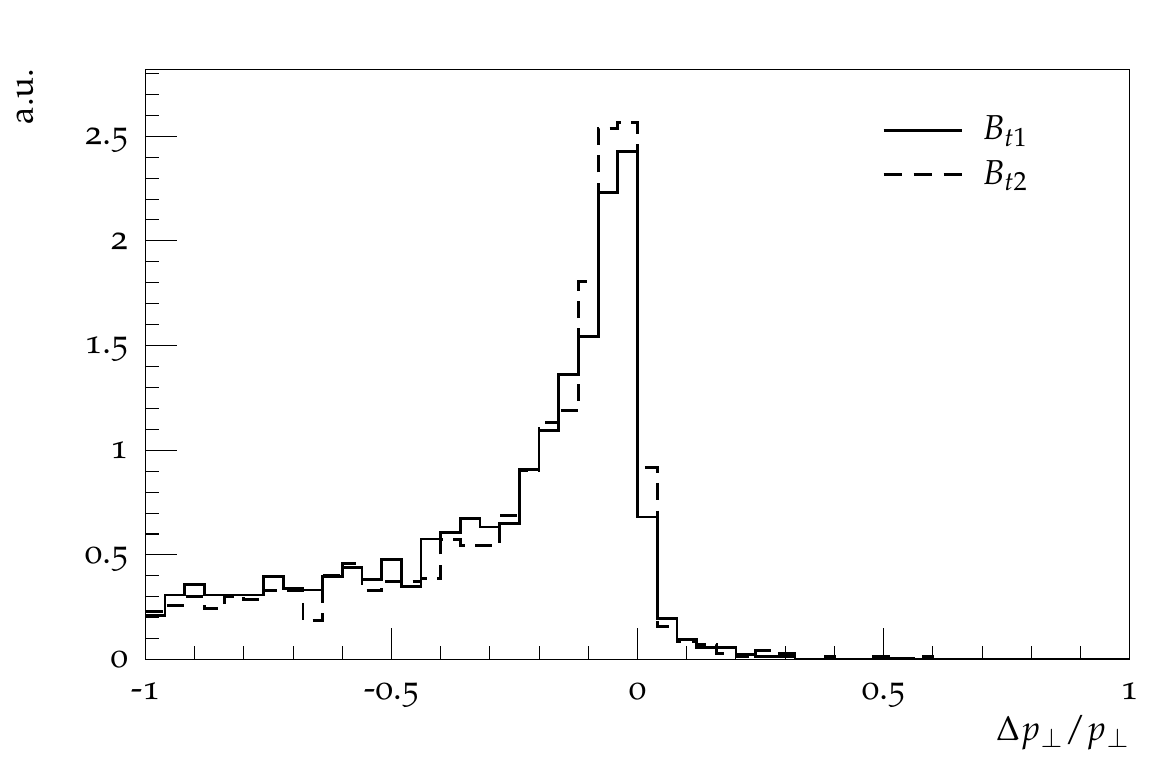}
 \includegraphics[width=0.32\textwidth]{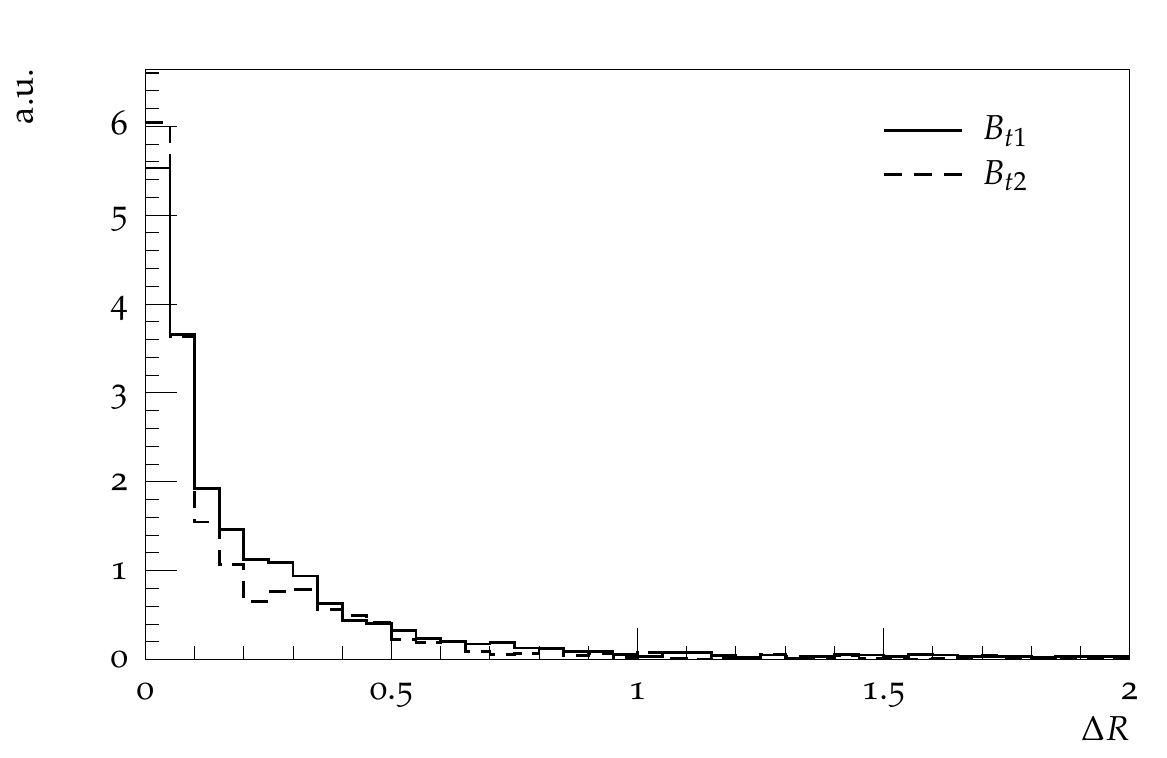}
 \end{center}
 \caption{Reconstruction quality of tagged top buckets of the leading $t\bar{t}$ background events in the MJ10 setup.}
 \label{fig:reco_quality_tt_mj10}
\end{figure}

We conclude this subsection with the reconstruction quality of tagged top buckets from the dominating $t\bar{t}$ SM background (in the MJ10 setup), which is shown in Fig.~\ref{fig:reco_quality_tt_mj10}.
Deviations from the MC truth partons are very small and our analysis setup is well suited for the background processes as well.
We observe that, although transverse momentum is reconstructed generally very accurately, the buckets tend to have lower values than the signal case (cf.~Fig.~\ref{fig:reco_quality_mj10}).
Final-state radiation off the boosted top quark may escape from the respective top bucket, and additional hard prongs from the matrix element that could lead to splash-in are not present.

\subsection{A note on the global metric}
\label{subsec:metric}

The reader might wonder whether the jets are not optimally assigned to the buckets due to the explicitly decoupled metric in Eqs.~(\ref{eq:bucket_global_metric}) and~(\ref{eq:bucket_decoupled_metric}).
This choice was made to reduce the combinatorial workload, but it is expected that the results 
does not change much even if a more democratic ordered metric, $0<\omega_{i+1}/\omega_i<1$, is used.
First, we observe that for any bucket the exchange of a jet with one from $B_\text{ISR}$ does not yield a lower metric by definition, independent of its weights. Secondly,
interchange of jets between two buckets ($B_i$ and $B_j$ with $\omega_i>\omega_j$) may lower the measure of $B_j$, but always at the cost of raising that of $B_i$. Because of the relative weight $\omega_i>\omega_j$, 
most of the interchanges are likely to increase the global measure. To find the global minimum, one has to consider a re-assignment of several jets simultaneously, the details of which depend on the specific weights chosen and is beyond the scope of this paper.
Note that, even if finite weights are used, the local minimum found with explicitly decoupled buckets gives an upper bound on $\Delta_\text{min}^2$, thus helping to reduce the huge number of permutations.

This argument is weakened if a metric is chosen that does not favour top buckets over Higgs buckets, i.e.~$\omega_1 , \omega_2 \ngtr \omega_3 , \omega_4$.
In our analysis, we chose to reconstruct top quarks from three prongs first, and only after that Higgs bosons from two prongs each.
This order reduces wrong assignments for both the signal and background.
Since SM processes containing Higgs bosons in the final state are rare, the mass distributions in the Higgs buckets can serve as side bands to experimentally determine the background cross-sections.

\section{Summary and outlook}
\label{sec:summary}

We have presented a novel approach to the very busy all-hadronic final state emerging from multiple heavy resonances, focusing on vectorlike top pair production at the LHC as the benchmark of our studies.
Since the standard techniques using large-radius fat jets suffer from splash-in contamination and jet overlap
in such a busy environment, in this paper we completely relied on separately resolved jets.
It was shown that this approach -- in combination with a bucket algorithm to reduce computational weight -- gives good results and can serve as an alternative channel in new physics searches, including a kinematic reconstruction of the vectorlike top mass.
The key ingredient is the mass-jump jet clustering algorithm, which is shown to greatly improve the performance compared to common jet algorithms.
This algorithm, which established the family of jet clustering with a terminating veto, is able to resolve nearby hard partons into separate jets, while it resembles common jet algorithms if the partons are well-isolated.
In addition to intrinsic jet properties, it introduces a dependence of the clustering history on 
two-jet properties, all formulated in terms of jet mass and mass ratios.
It is this flexibility that outputs jets with variable effective radii, which leads to superior results compared to the fixed-radius variants.

While a $\chi^2$-like measure could give a more accurate assignment of the jets to the various buckets, we gave an argument that the difference to our computationally inexpensive ansatz is not expected to be large.
Another possible improvement is to require a certain number of $b$-tagged jets for each top and Higgs candidate.
We did not include this option in our analysis because bottom tagging is difficult in such busy final states, and also it would require matching between tagged jets and mass-jump jets, which have not yet been investigated by the experimental collaborations.
Our results give a conservative estimate in this respect.

On top of the phenomenological study of vectorlike top pair production, we investigated the quality of reconstruction of the top quarks and Higgs bosons.
Whereas the majority of tagging algorithms for boosted resonances assumes their isolation, we showed that our approach performs excellently in identifying the correct jet combinations even in this very busy and unclean environment.
This  study enters uncharted and often neglected territory when it comes to taggers, yet the results are promising 
and we expect that jet clustering algorithms with a terminating veto will find their place in future studies of high-multiplicity processes.
The algorithm dubbed \texttt{ClusteringVetoPlugin} is publicly available in the \fastjet contributions package~\cite{fastjet-contrib}.

\begin{acknowledgments}

M.S. is thankful to Michihisa Takeuchi for helpful discussion and advice.
We are also grateful to Gavin Salam of the \fastjet team for useful comments on and support with the \fastjet plugin.
The work of S.P.L. and M.S. was supported in part by the Program for Leading Graduate Schools, MEXT, Japan.
The work of S.P.L. was also supported in part by a JSPS Research Fellowships for Young Scientists.
This work was supported by Grant-in-Aid for Scientific research No. 26104001, 26104009, 26247038, 26800123, and also by World Premier International Research Center Initiative (WPI Initiative), MEXT, Japan.

\end{acknowledgments}

%% --------------------                                                         
%% |   Bibliography   |                                                         
%% --------------------                                                         
%%
\input{inspire_bibliography}

% replace manually
% \bibitem{fastjet-contrib}
%   \url{http://fastjet.hepforge.org/contrib/}

\end{document}

%% file: inspire_bibliography.tex
\providecommand{\href}[2]{#2}\begingroup\raggedright\endgroup